\documentclass[apj]{emulateapj}
\usepackage{comment}
\usepackage{ifthen}
\usepackage{lineno}
\usepackage{placeins}
\usepackage{amsmath}
\usepackage{xspace}
\usepackage{hyperref}
\usepackage{array}

\newcommand{\ctbd}[1]{}

\newcommand{\thislambda}{\ensuremath{-10.9_{-3.8}^{+3.6}}}

\newcommand{\teff}{$T_\text{eff}$\xspace}    
\newcommand{\logg}{$\log g$\xspace}
\newcommand{\feh}{[Fe/H]\xspace}

\newboolean{emulateapj}
\setboolean{emulateapj}{true}

\newboolean{rvtablelong}
\setboolean{rvtablelong}{true}

\newboolean{astroph}
\setboolean{astroph}{true}

\shortauthors{Zhou et al. }
\shorttitle{Obliquities of planets around HD 106315}
\ifthenelse{\boolean{emulateapj}}{
    \newcommand{\titledag}{$\dagger$}
}{
    \newcommand{\titledag}{\dagger}
}
\ifthenelse{\boolean{emulateapj}}{
    \newcommand{\titlestar}{$\star$}
}{
    \newcommand{\titlestar}{\star}
}
\ifthenelse{\boolean{emulateapj}}{
    
}{
    
}

\begin{document}


\title{The warm Neptunes around HD 106315 have low stellar obliquities}

\author{
  George Zhou\altaffilmark{1,\titledag},
  Joseph E. Rodriguez\altaffilmark{1},
  Andrew Vanderburg\altaffilmark{2,\titlestar},
  Samuel N. Quinn\altaffilmark{1},
  Jonathan Irwin\altaffilmark{1},
  Chelsea X. Huang\altaffilmark{3},
  David W. Latham\altaffilmark{1},
  Allyson Bieryla\altaffilmark{1},
  Gilbert A. Esquerdo\altaffilmark{1},
  Perry Berlind\altaffilmark{1}, and
  Michael L. Calkins\altaffilmark{1}
}

\altaffiltext{1}{Harvard-Smithsonian Center for Astrophysics, 60 Garden Street, Cambridge, MA 02138 USA; george.zhou@cfa.harvard.edu}
\altaffiltext{2}{Department of Astronomy, The University of Texas at Austin, 2515 Speedway, Stop C1400, Austin, TX 78712}
\altaffiltext{3}{Department of Physics, and Kavli Institute for Astrophysics and Space Research, M.I.T., Cambridge, MA 02139, USA}

\altaffiltext{\titledag}{Hubble Fellow}
\altaffiltext{\titlestar}{Sagan Fellow}


\begin{abstract}

We present the obliquity of the warm Neptune HD 106315c measured via a series of spectroscopic transit observations. HD 106315c is a $4.4\,R_\mathrm{Earth}$ warm Neptune orbiting a moderately rotating late F-star with a period of 21.05\,days. HD 106315 also hosts a $2.5\,R_\mathrm{Earth}$ super-Earth on a 9.55\,day orbit. Our Doppler tomographic analyses of four transits observed by the Magellan/MIKE, HARPS, and TRES facilities find HD 106315c to be in a low stellar obliquity orbit, consistent with being well aligned with the spin axis of the host star at $\lambda = \thislambda{} ^\circ$. We suggest, via dynamical N-body simulations, that the two planets in the system must be co-planar, and thus are both well aligned with the host star. HD 106315 is only the fourth warm Neptune system with obliquities measured. All warm Neptune systems have been found in well aligned geometries, consistent with the interpretation that these systems are formed in-situ in the inner protoplanetary disk, and also consistent with the majority of \emph{Kepler} multi-planet systems that are in low obliquity orbits. With a transit depth of 1.02 mmag, HD 106315c is among the smallest planets to have been detected in transit spectroscopy, and we discuss its detection in the context of \emph{TESS} and the next generations of spectrographs.


\setcounter{footnote}{0}
\end{abstract}



\section{Introduction}
\label{sec:introduction}

The \emph{Kepler} mission unveiled a population of super Earths and Neptunes in periods of $10-200$ days that occupy more than 50\% of sun-like stars \citep[e.g.][]{2013ApJ...766...81F,2013ApJ...778...53D}. With no analogues in our own Solar System, we are beginning to question the peculiarities of our own existence \citep[e.g.][]{2015PNAS..112.4214B}. We now think that some of these planets were potentially formed in-situ in the inner protoplanetary disk. In this scenario, the rocky cores of these planets may have formed at distances of $\sim 1$\,AU, and migrated inward before the dissipation of the inner disk to accrete a substantial $(\gtrsim 10\%)$ gaseous envelope \citep[e.g.][]{2014ApJ...797...95L,2016ApJ...817L..17B}, forming the systems of super Earths and Neptunes we observe. 

The system obliquity -- the angle between the orbit normal and the spin-axis of the host star, is a tracer for the early history of close-in planets. In particular, the obliquities of hot Jupiters exhibit a wide distribution of angles \citep[especially for those not thought to be influenced by star-planet tidal interactions, e.g.][]{2010ApJ...718L.145W,2012ApJ...757...18A}, suggesting that some of these planets were thrown inward following earlier dynamical episodes. 

Many hot Neptunes in close-in orbits have been found to be in highly misaligned orbits. Transit spectroscopic and star-spot crossing observations showed HAT-P-11b to be in a 4.9\,day polar orbit about a K4V star \citep{2010ApJ...710.1724B,2010ApJ...723L.223W,2011PASJ...63S.531H,2011ApJ...743...61S}; the lack of consecutive spot crossing events in the 5.7\,day period super Neptune WASP-107b system led \citet{2017AJ....153..205D} to infer that the planet must reside in a misaligned orbit; similarly, recent spectroscopic observations of the 2.5\,day period hot Neptune GJ 436b found it to also orbit in a polar geometry \citep{2018Natur.553..477B}. 

In contrast, the systems of warm Neptunes we have surveyed so far are all well aligned to the spins of their host stars. Spectroscopic transit observations by \citet{2013ApJ...771...11A} found Kepler-25c, a 4.48\,$R_\mathrm{Earth}$ 12\,day period Neptune to be well aligned with its F-star host. Additional measurements of the stellar spin inclination via asteroseismology found that the host star Kepler-25 is not inclined with our line of sight \citep{2014PASJ...66...94B,2016ApJ...819...85C}, consistent with a low obliquity plane for both planets in the system. Further asteroseismology analyses of the warm Neptune systems Kepler-50 and Kepler-65 found them both to be consistent with the aligned geometry as well \citep{2013ApJ...766..101C}.

Is this the start of a dichotomy distinguishing warm Neptune systems and single close-in Neptunes in obliquity? Measuring the obliquities of transiting Neptunes is a challenging endeavor given that the transit depth is shallower by an order of magnitude compared to equivalent Jovian systems. While the \emph{Kepler} mission found numerous warm Neptune systems, few orbit bright enough stars for further examination of their obliquities. With photometry from the \emph{K2} survey, the F star HD 106315 was identified to host two planets, a $2.51\pm 0.12 \,R_\mathrm{Earth}$, $12.6\pm3.2\,M_\mathrm{Earth}$ super Earth at 9.55\,day period, and a $4.31_{-0.27}^{+0.24}\,R_\mathrm{Earth}$, $15.2\pm3.7 M_\mathrm{Earth}$ Neptune at 21.06\,day period \citep{2017AJ....153..256R,2017AJ....153..255C,2017A&A...608A..25B}. Key system parameters from literature are listed in Table~\ref{tab:literature_properties}. The brightness ($V_\mathrm{mag} = 9.004$) and rotation ($v\sin i_\mathrm{rot} = 12.9 \pm 0.4\,\mathrm{km\,s}^{-1}$) of HD 106315 makes the system ideal for follow-up observations. In this paper, we present spectroscopic observations of the transits of HD 106315c, gathered from three independent facilities, covering four events, to measure the projected obliquity of the system.

\begin{table}
    \centering
  
  \caption{\label{tab:literature_properties}Key literature properties of HD 106315 system$^a$}
 {\renewcommand{\arraystretch}{2.0}
 \begin{tabular}{lr}
    \hline\hline
    RA &  12:13:53.394 \\
    DEC &  -00:23:36.54 \\
    $V_\text{mag}$ &  9.004 \\
    $M_\star \, (M_\odot)$ &  $1.027_{-0.029}^{+0.034}$  \\
    $R_\star \, (R_\odot)$ &  $1.281_{-0.058}^{+0.051}$ \\
    $T_\text{eff}\,(\text{K})$  & $6254_{-51}^{+55}$ \\
    $v\sin i_\mathrm{rot}\,(\text{km\,s}^{-1})$  & $12.9\pm 0.4$ \\
    Parallax (mas) & $9.1163\pm 0.0569$ \\
    \hline
    \emph{Planet b} &\\
    Period (days)  & $9.55385_{-0.00072}^{+0.00095}$ \\
    $M_p\,(M_\text{Earth})$  & $12.6 \pm 3.2$  \\
    $R_p\,(R_\text{Earth})$  & $2.51 \pm 0.12$ \\
    
    \hline
    \emph{Planet c} &\\
    Period (days)& $21.0580\pm0.0022$\\
    $M_p\,(M_\text{Earth})$ & $15.2 \pm 3.7$\\
    $R_p\,(R_\text{Earth})$ & $4.31_{-0.27}^{+0.24}$\\
    \hline
    \hline
  \end{tabular}}

\footnotesize $^{a}$All parameters from \citet{2017AJ....153..256R} except for planet masses from \citet{2017A&A...608A..25B} and parallax from \emph{Gaia} DR2.
\end{table}

\section{Spectroscopic observations and analysis}
\label{sec:obs_and_analyses}

Time series spectroscopic observations during transit can yield the projected obliquities of the orbiting planet. During the transit, the planet occults the rotating stellar surface, successively blocking different velocity components of the stellar line profile. To map this spectroscopic transit, we monitor for changes in the line profile for signatures of this transiting shadow  \citep{1924ApJ....60...15R,1924ApJ....60...22M,2010MNRAS.403..151C,2010MNRAS.407..507C}. By modeling the track of the planetary shadow over the course of a transit, we can infer the projected angle between the orbit normal and the stellar spin axis. Spectroscopic transits of planets have yielded the projected obliquities of more than 100 planetary systems to date\footnote{TEPcat \url{http://www.astro.keele.ac.uk/jkt/tepcat/obliquity.html}}. 

\begin{figure*}
    \centering
    \includegraphics[width=18cm]{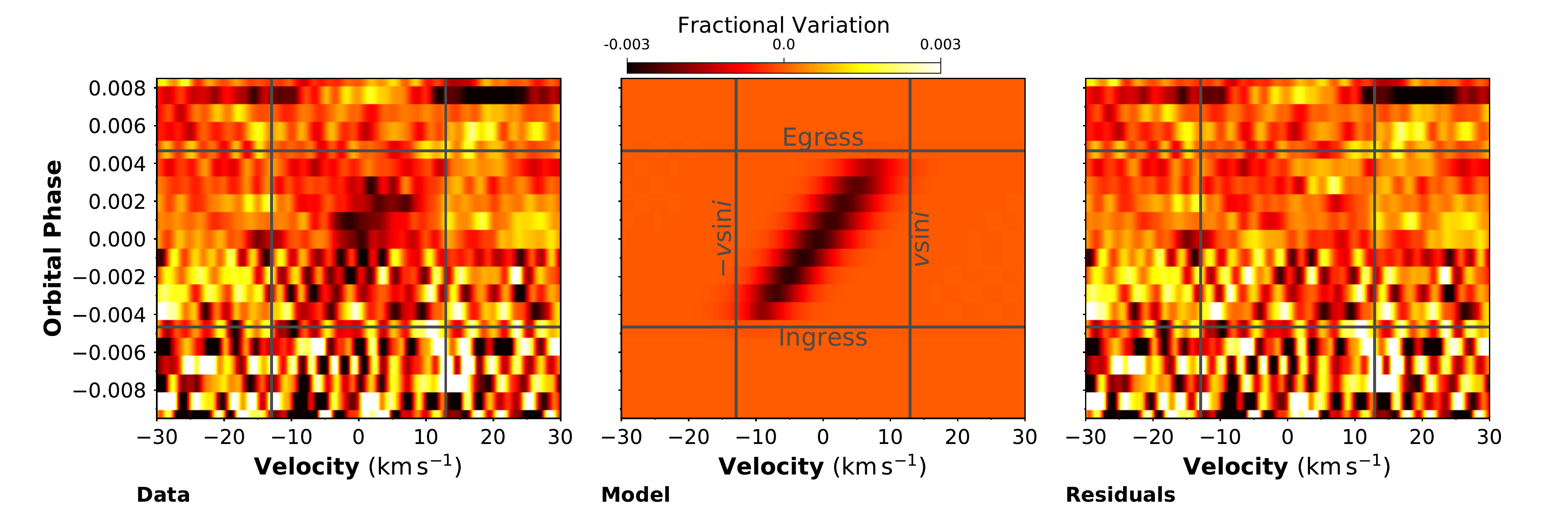}
    \caption{Doppler tomographic maps showing the combined planetary transit signal from four transits obtained with MIKE, HARPS, and TRES. The transiting planet successively blocks parts of the rotating stellar surface, inducing a `shadow' in the stellar line profile. Doppler tomographic maps show these line profile variations as a function of stellar rotational velocity (x-axis) and orbital phase (y-axis). The \textbf{left panel} shows the combined observations, temporally binned to 0.0005 in phase (15 minutes), with the planetary transit signal seen as the dark trail crossing from bottom left to top right. The vertical lines mark the $v\sin i_\mathrm{rot}$ of the star, and the horizontal lines mark the timings of ingress and egress. The  \textbf{central panel} shows the best fit model to the detected planetary transit trail, and the \textbf{right panel} shows the residuals after the model is subtracted. Note the signal-to-noise improves after mid-transit due to the partial transit coverage of the MIKE observation.}
    \label{fig:DT_combined}
\end{figure*}

The detectability of the spectroscopic transit is directly proportional to the transit depth induced by the planet, and the projected rotational broadening of the star. As such, with a depth of 0.102\%, the transit HD 106315c is the shallowest of all successful spectroscopic transit measurements to date. The following sections describe the four transit observations gathered from the MIKE/Magellan, HARPS, and TRES facilities that were used in our analyses. Figure~\ref{fig:DT_combined} shows the combined Doppler tomographic map of these observations, yielding a detection of the planetary transit. 

\subsection{Spectroscopic transit of HD 106315c from MIKE}
\label{sec:mike}

We observed a partial spectroscopic transit of HD 106315c on 2018 Mar 01 with the Magellan Inamori Kyocera Echelle \citep[MIKE,][]{2003SPIE.4841.1694B} on the 6.5\,m Magellan Clay telescope, located at Las Campanas Observatory, Chile. To achieve the highest spectroscopic resolution possible, the observations were performed with the $0.35\arcsec$ slit, yielding a resolution of $\lambda / \Delta \lambda \equiv R=65000$ in the red arm of the spectrograph ($4900-10000$ \,\AA) and $R=85000$ in the blue arm ($3200-5000$\,\AA). A total of 81 observations were made over the time period of 03:03 -- 09:08 UTC (covering orbital phases 0.997 to 1.009, with the transit being from phase 0.9953 to 1.0047), with the target staying below airmass 1.7, and seeing of $\sim 1\arcsec$, throughout the observations. Each observation was a 200\,s integration in the blue and red cameras. Wavelength calibration is provided via ThAr arc lamp exposures taken every $\sim 30$\,minutes. Flat field calibrations were obtained in the afternoon via a quartz lamp and a diffuser. The observations were reduced using the Carnegie \emph{CarPy} package \citep{2000ApJ...531..159K,2003PASP..115..688K}.

To measure the planetary transit signal, we derive stellar line profiles for each observation via a least-squares deconvolution technique against a synthetic spectral template \citep{1997MNRAS.291..658D}. A non-rotating infinite resolution template matching the spectral classification of HD 106315 is synthesized via the \emph{SPECTRUM} code \citep{1994AJ....107..742G} using the ATLAS9 atmosphere models \citep{2004astro.ph..5087C}. The stellar line profile, derived using the deconvolution technique, is the kernel that convolves the synthetic template to the observation, and contains contributions from the stellar rotational broadening, non-rotating broadening effects such as macroturbulence, and broadening due to the instrument profile. To derive a high signal-to-noise broadening profile for each spectrum, we perform deconvolutions over individual orders spanning $4900-6200$\,\AA{} in wavelength, and then weighted-average the resulting line profile kernels based on the signal-to-noise of the derived line profiles from each order.

\subsubsection{Correction of instrument profile variations via telluric features}
\label{sec:telluric_cor}

\begin{figure*}
    \centering
    \includegraphics[width=12cm]{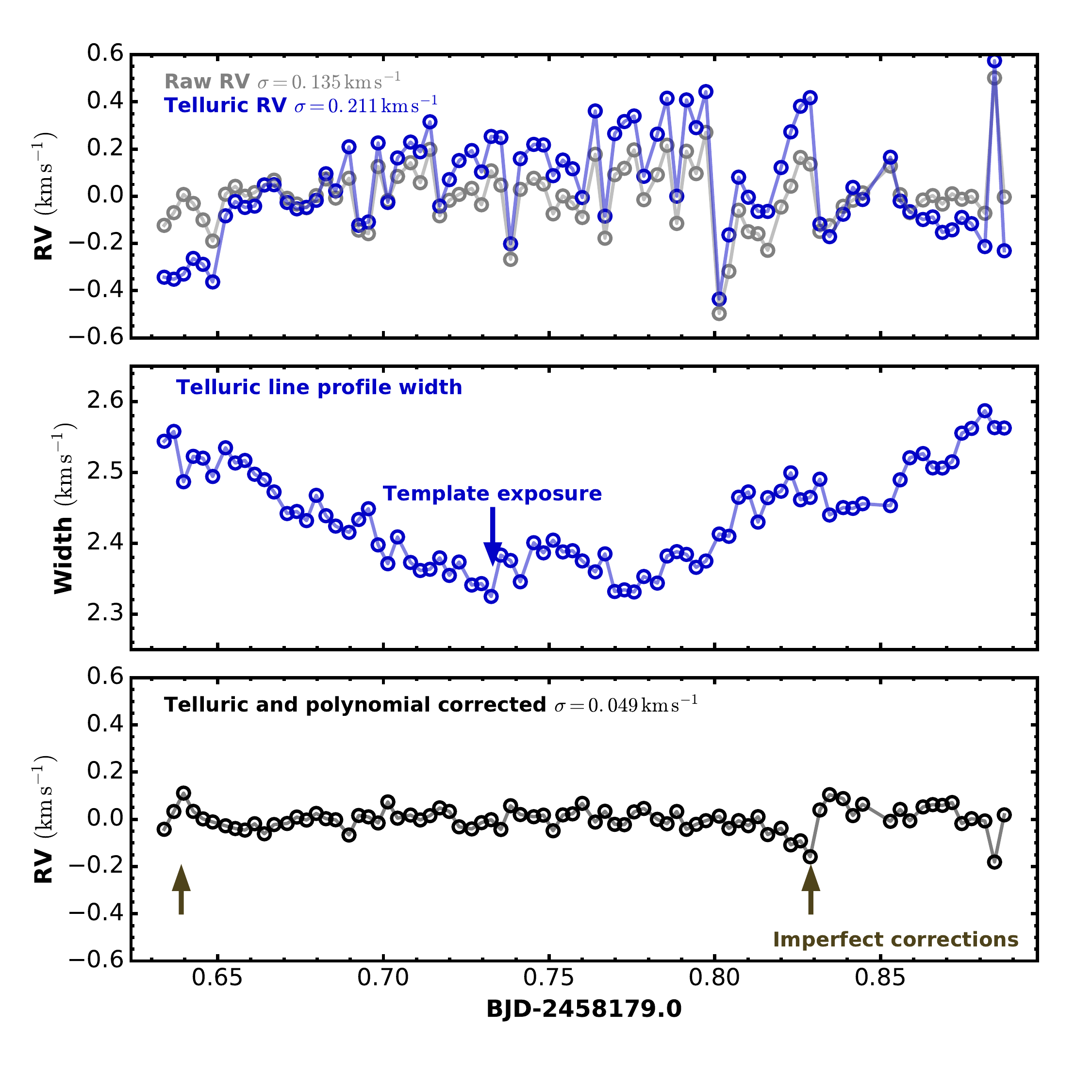}
    \caption{Radial velocities of the MIKE observations on 2018 Mar 01. The \textbf{top panel} shows the velocities derived from the uncorrected stellar line profiles (grey) and telluric features (blue). The point-to-point scatter of the velocities are $\sim 0.135\,\mathrm{km\,s}^{-1}$, and are indicated in the figure labels. The \textbf{central panel} shows variations in the telluric feature widths over the course of the night. The template exposure used to correct for the slit-based spectral variations is marked in the figure. The \textbf{bottom panel} shows the velocities derived from the stellar line profiles after they were corrected using the telluric feature-derived instrument profile. The resulting point-to-point scatter is $\sim 0.05\,\mathrm{km\,s}^{-1}$. Note that despite these corrections, the velocities systematically deviate from zero at multiple locations, indicating that our telluric-derived line profile does not fully model variations in the instrument profile. }
    \label{fig:RV_magellan}
\end{figure*}

Since MIKE is a slit-fed spectrograph, the instrumental line spread function is subject to variations induced by the target star's position on the slit. As the star `wobbles' on the slit due to variable seeing and guiding errors, the spectroscopic instrument profile changes in shape and velocity. In addition, thermal and pressure fluctuations can also induce changes in the instrument profile. These variations traditionally prohibit slit-based spectrographs from achieving radial velocity precision of better than $\sim 200\,\mathrm{m\,s}^{-1}$ without additional corrections (Figure~\ref{fig:RV_magellan}), and in our case completely wash out the spectroscopic signal of a 1\,mmag planetary transit. Similar instrument profile variations were discussed in \citet{2013ApJ...771...11A} regarding the Doppler tomographic detection of Kepler-25c. To correct for these variations, we take inspiration from early precise radial velocity spectroscopists and make use of the telluric water features to correct for the instrument profile variations \citep{1973MNRAS.162..243G,1973MNRAS.162..255G}. 

\begin{figure*}
    \centering
    \begin{tabular}{ccc}
        \textbf{MIKE Red arm} &&\\
         \includegraphics[width=5cm]{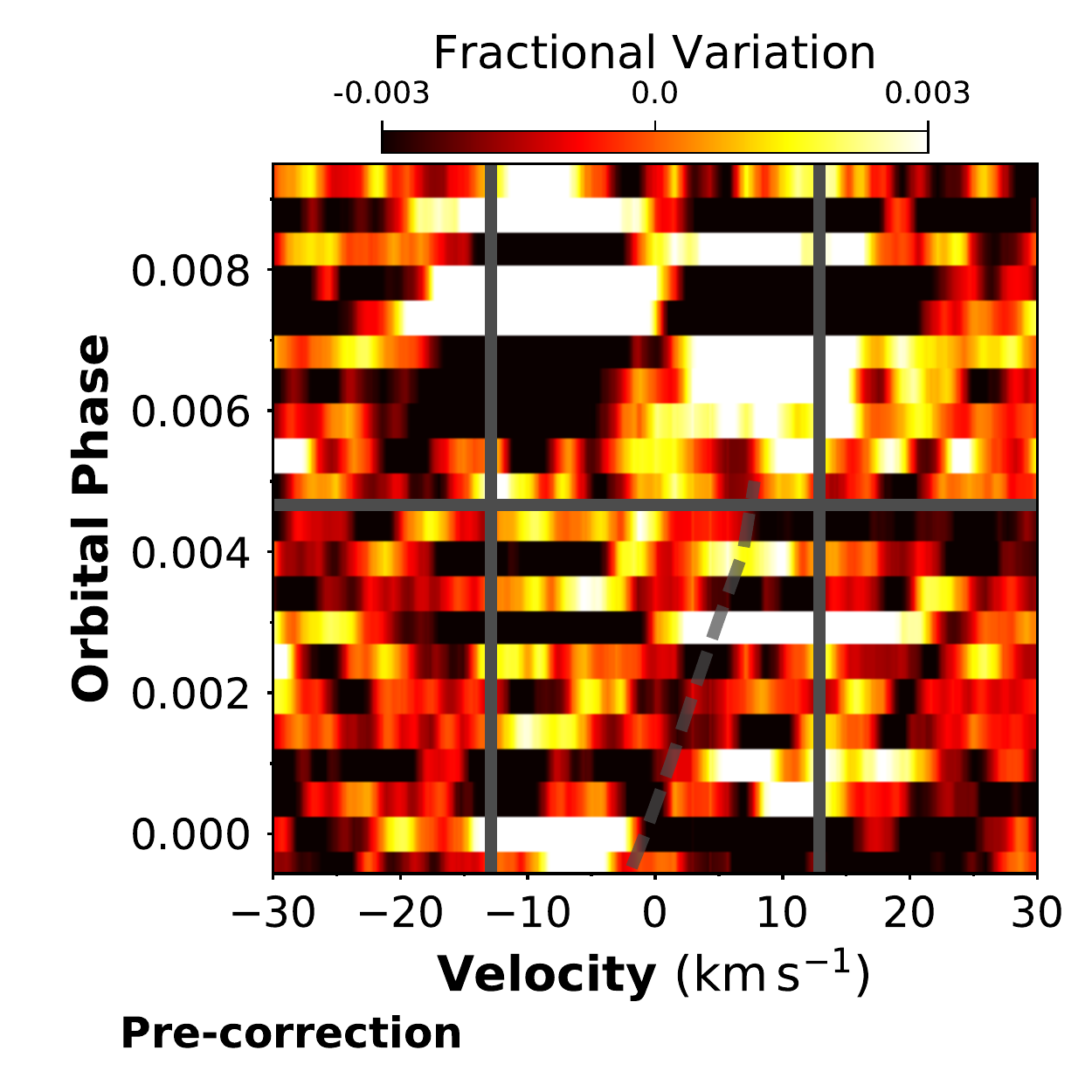} &
         \includegraphics[width=5cm]{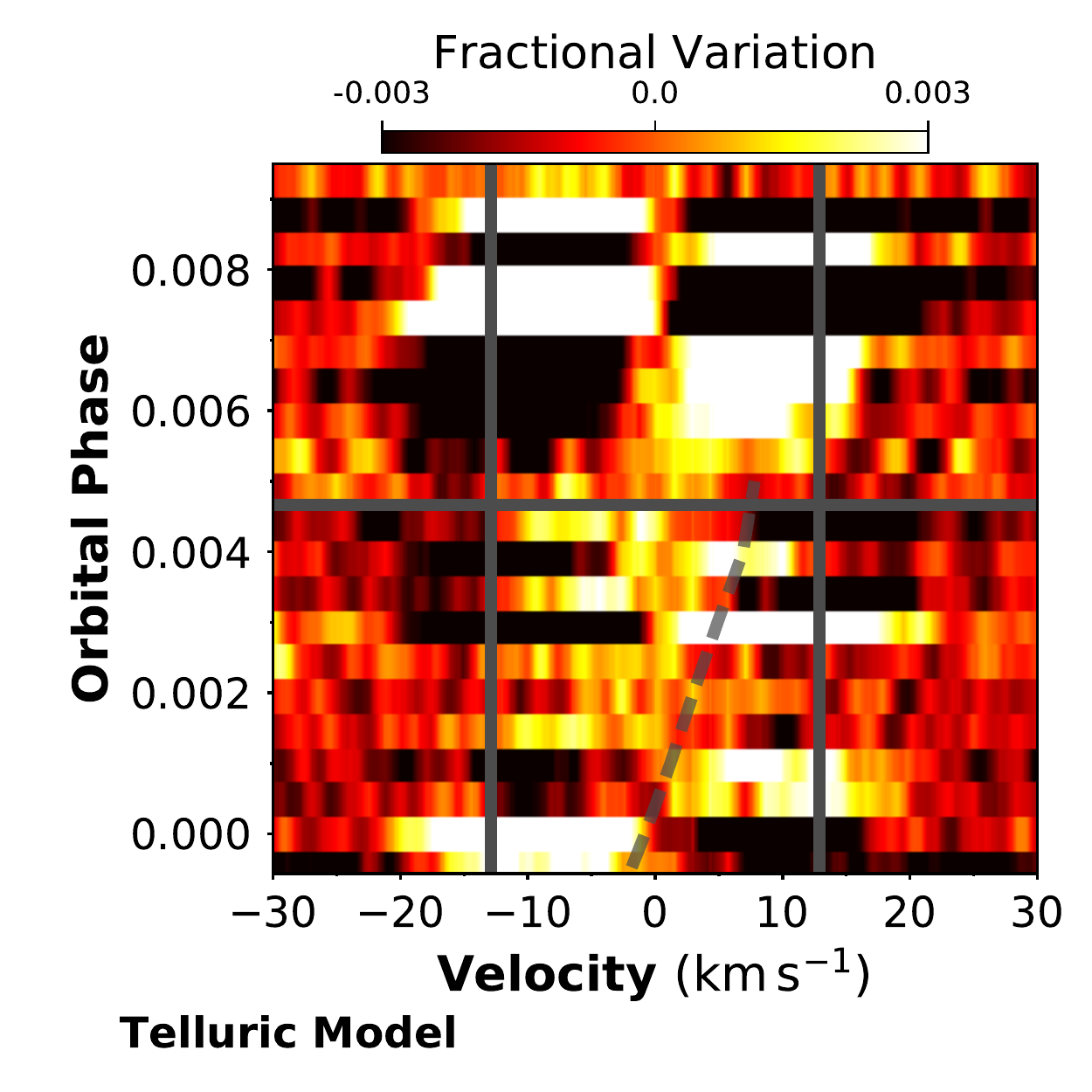} &
         \includegraphics[width=5cm]{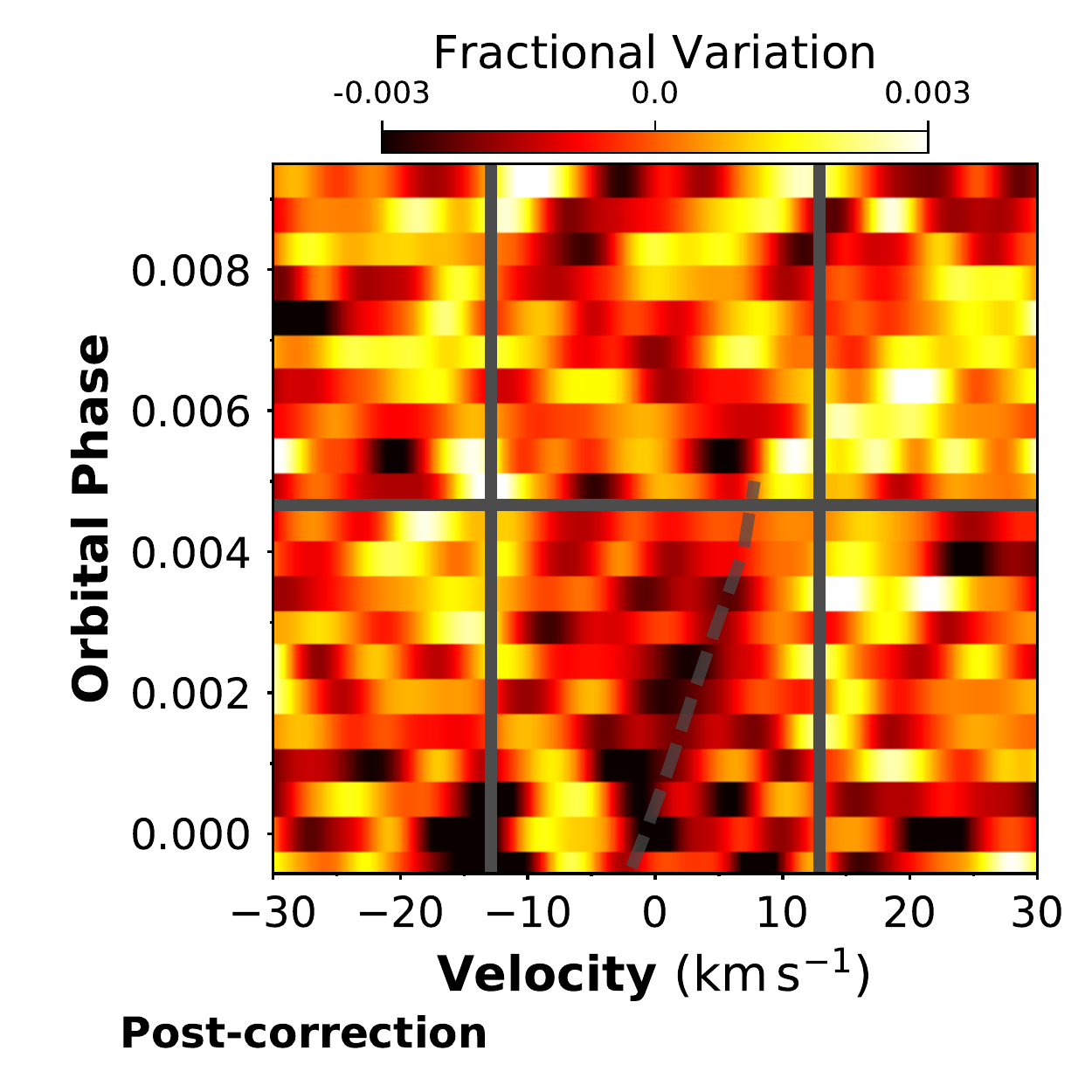} \\
        \textbf{MIKE Blue arm} &&\\
         \includegraphics[width=5cm]{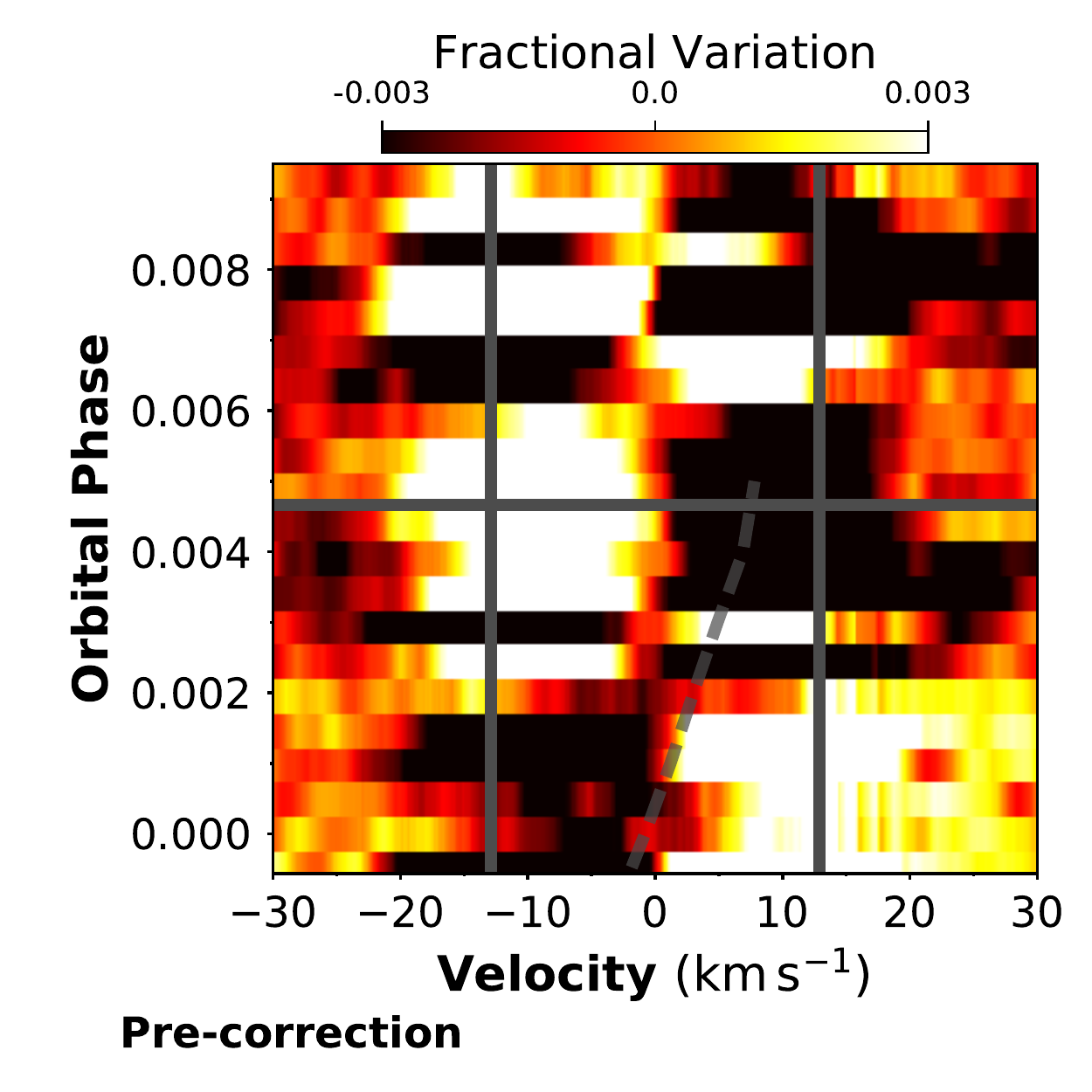} &
         \includegraphics[width=5cm]{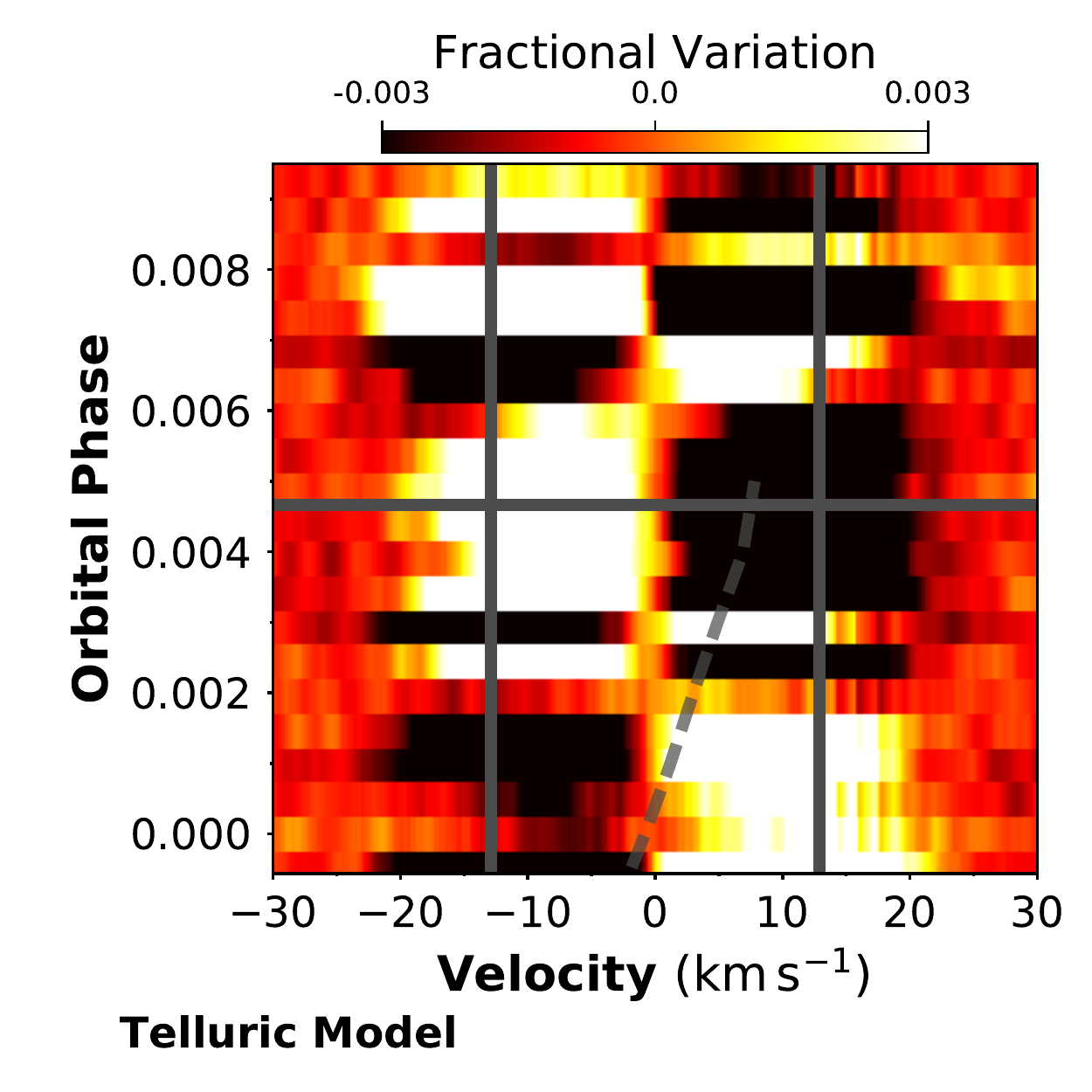} &
         \includegraphics[width=5cm]{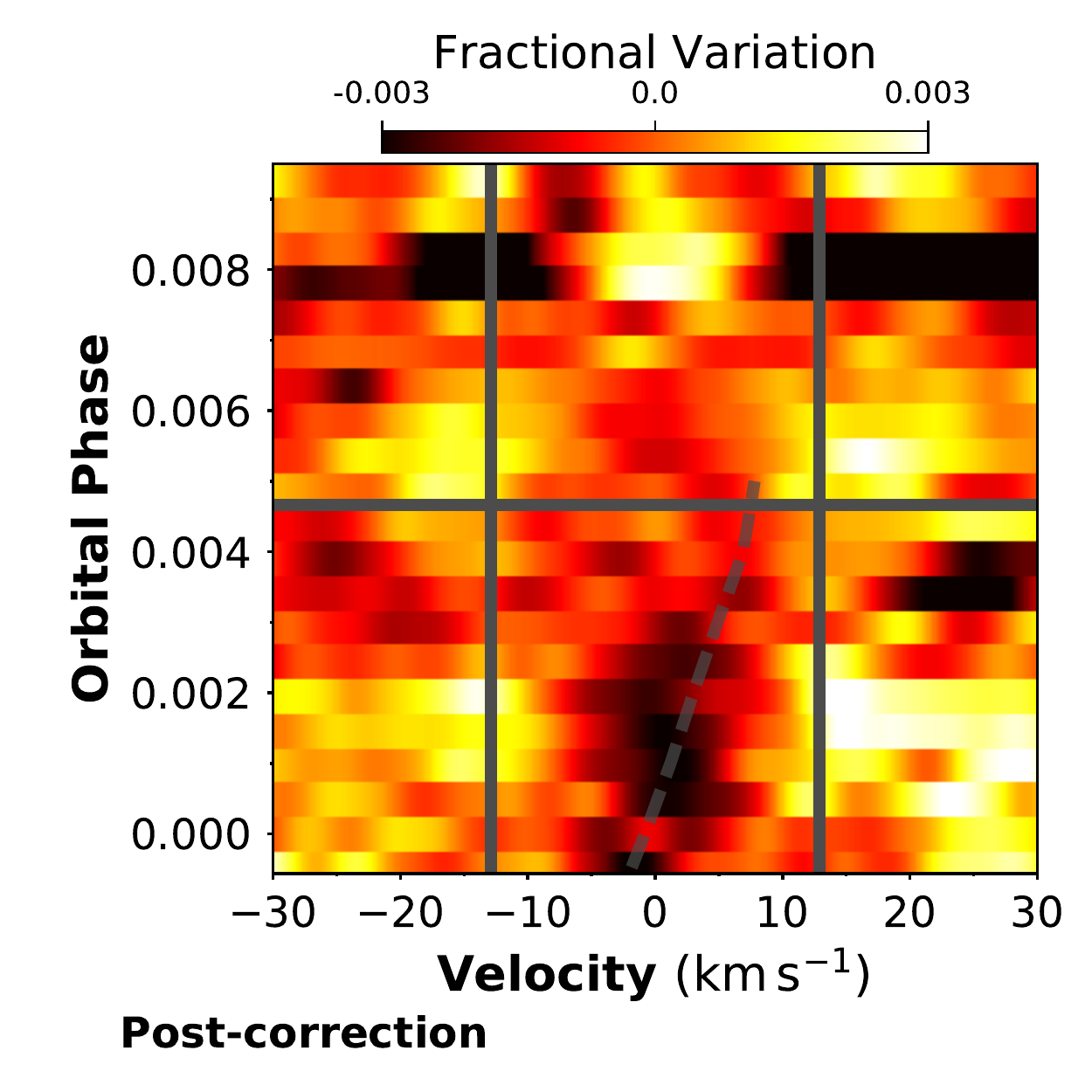} \\         
    \end{tabular}
    \caption{Doppler tomographic maps showing the stellar line profile variations observed by MIKE during the partial transit on 2018 Mar 01, phase binned at intervals of 0.0005 (15 minutes, $\sim 5$ exposures). The observations span from mid-transit to post-egress. The \textbf{left panels} show the uncorrected stellar line profile variations. The \textbf{central panels} show the modeled instrument profile corrections derived from variations in the telluric water features. The \textbf{right panels} show the corrected stellar line profiles. The \textbf{top panels} show the observations from the red arm of MIKE, the \textbf{bottom panels} show the blue arm observations. Though the corrections are not perfect, we can distinguish the planetary transit through the centre of the stellar line profile in the corrected dataset. The modeled velocity of the planetary shadow is marked by the dashed line to aid the eye. Imperfections exist between the telluric model and the observed stellar profile. These imperfections, seen as red noise affecting the dataset, are at the same location as the velocity offsets marked in Figure~\ref{fig:RV_magellan}.}
    \label{fig:mikeDT}
\end{figure*}

To correct for instrument profile variations during the transit, we search for a kernel that convolves the instrument profile of each observation to that of a template spectrum. The instrument profile of each observation is assumed to be the line profile of the telluric absorption features, derived from a least-squares deconvolution over the unsaturated telluric water features at $7900-8100$\,\AA{} and $7000-7100$\,\AA{} against an artificial $R=10^6$ template synthesized with the \emph{ESO SkyCalc Sky Model Calculator} \citep{2014A&A...568A...9M}. We then choose the broadest instrument profile during the night as the template, and fit for an asymmetric Gaussian kernel that would convolve the instrument profile of each observation to that of the template. The asymmetric Gaussian kernel is parameterized by two width terms $(\sigma_\mathrm{left},\sigma_\mathrm{right})$, a height scaling factor, and a velocity offset. The kernel parameters are fitted for with a Markov chain Monte Carlo process, using the module \emph{emcee} \citep{2013PASP..125..306F}. At each iteration, the trialed kernel is convolved with the telluric profile, and the resulting `corrected' profile is compared against the template profile, with the goodness of fit used to maximize the likelihood. The best fit asymmetric Gaussian kernel is then convolved with the stellar broadening kernel. This correction removes the large point-to-point `jumps' in the instrument line profile and associated velocities, but a smoothly varying trend remains, possibly due to intrinsic variations in the telluric lines over the night as a function of water content and observational airmass. Since we do not expect to observe any velocity variation during the transit within our detection limits, we shift each observed stellar line profile to rest by fitting for their velocity variations with a fourth-order polynomial. The telluric profiles and corrections are derived using spectra from red arm of MIKE, and the same corrections are applied to spectra from both the blue and red arms. The velocities derived from the stellar line profiles before and after the telluric and polynomial corrections are plotted in (Figure~\ref{fig:RV_magellan}), and the line profile variations, as well as the telluric correction, are shown in Figure~\ref{fig:mikeDT}. We note that at times the telluric lines do not represent the instrument profile well, and as such the velocity and line profile corrections we apply exhibit systematic offsets. These correction imperfections are labelled in Figure~\ref{fig:RV_magellan}, and correspond with poorly corrected line profile residuals in Figure~\ref{fig:mikeDT}, near phase 0 and phase 0.008. 

\subsection{Spectroscopic transit of HD 106315c from TRES}
\label{sec:tres}

One full transit of HD 106315c was obtained with the Tillinghast Reflector Echelle Spectrograph \citep[TRES][]{Furesz2014}, on the 1.5\,m telescope at the Fred Lawrence Whipple Observatory, Arizona, USA. TRES is a fiber fed echelle spectrograph with a wavelength coverage of $3900-9100$\,\AA, spanning 51 echelle orders at a resolution of $R=44000$. A total of 36 observations were obtained on 2017 Apr 20. Each observation consists of three consecutive exposures, totalling 450\,s integration time, combined to reduce the impact of cosmic rays. Wavelength solutions are provided by ThAr lamp exposures that bracket each observation. The observations are reduced as per \citet{2010ApJ...720.1118B}.

\begin{figure}
    \centering
    \includegraphics[width=7cm]{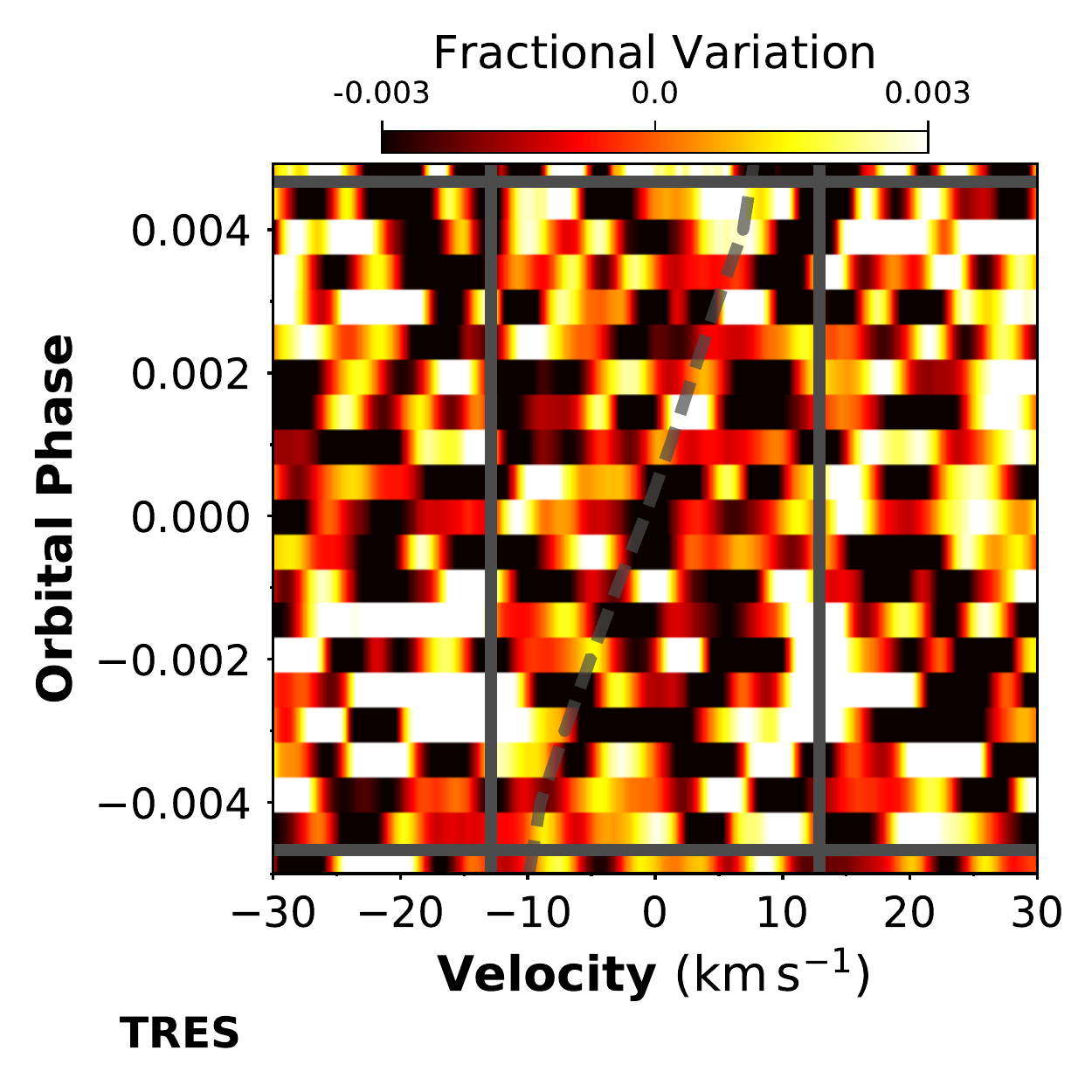}
    \caption{Doppler tomographic map showing the line profile residuals from the TRES observations obtained on 2017 Apr 20, phase binned at intervals of 0.0005 (15 minutes, 2 exposures). A partial transit was obtained, starting just before ingress, and ending before egress occurs. The figure follows the format introduced in Figure~\ref{fig:DT_combined}, the modeled velocity of the planetary shadow is marked by the dashed line to aid the eye.}
    \label{fig:tresDT}
\end{figure}

The line profile analysis of the TRES transit largely follows the process described in \citet{2016MNRAS.460.3376Z}. Line profiles are derived from order-by-order least-squares deconvolutions over 34 orders spanning the spectral range $3900-6250$\,\AA. The spectral template used in the convolution is the same as that used for the MIKE analyses, generated with the ATLAS9 non-rotating synthetic templates. TRES is a fiber-fed echelle spectrograph typically capable of measuring radial velocities at the $10\,\mathrm{m\,s}^{-1}$ level \citep{2014ApJ...787...27Q}, and as such no telluric corrections were necessary in the processing of the TRES observations. The resulting line profile variations measured by TRES are shown in Figure~\ref{fig:tresDT}.

In addition to the full transit observed on 2017 Apr 20, 30 other TRES observations were obtained of HD 106315. These included series of observations taken near transit on 2017-03-09 that were obtained in poor conditions, and not used in the Doppler tomographic analyses, as well as a set of spectra taken out of transit presented in \citet{2017AJ....153..256R}. These observations are used in the line broadening profile analyses described in Section~\ref{sec:linebroad}.

\subsection{Archival spectroscopic transits of HD 106315c from HARPS}
\label{sec:harps}

Transits of HD 106315c were observed with the High Accuracy Radial Velocity Planet Searcher \citep[HARPS][]{2003Msngr.114...20M} on the ESO 3.6\,m telescope at La Silla Observatory, Chile. HARPS is a fiber-fed echelle spectrograph with a spectral resolution of $R=115000$, and a wavelength coverage of $3780 - 6910$\,\AA. Two transits were observed with HARPS: one on 2017 March 9 consisting of 75 exposures with 350s integration times, and one on 2017 March 30, consisting of 47 exposures with 500s integration times. On both nights the observations covered the entire transit events.  The reduced 2D echelle spectra were obtained from the public ESO archive. 

\begin{figure}
    \centering
    \includegraphics[width=7cm]{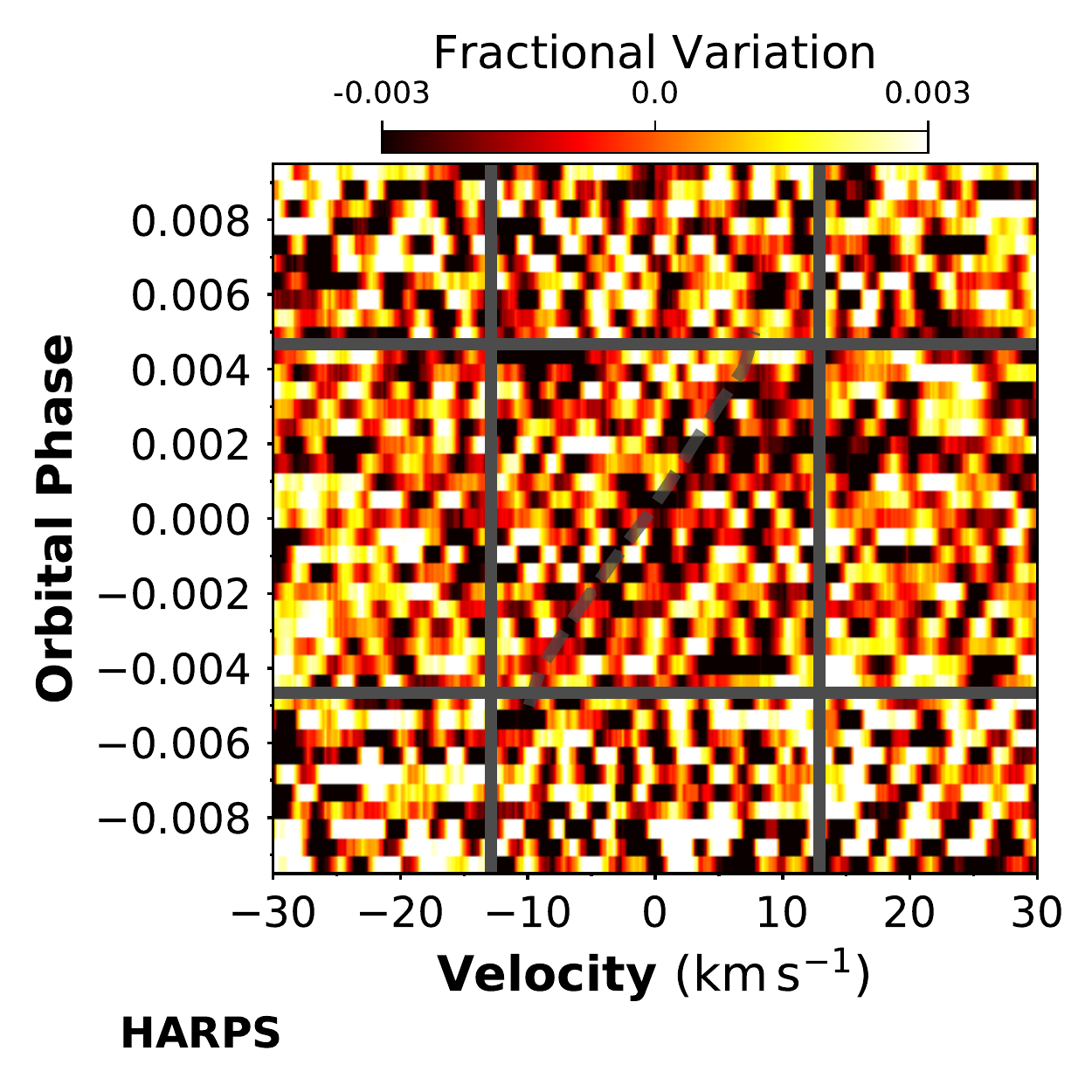}
    \caption{Doppler tomographic map showing line profile variations from the HARPS observations on 2017 Mar 09 and 2017 Mar 30, phase binned at intervals of 0.0005 (15 minutes, $\sim 4$ exposures). The figure follows the format introduced in Figure~\ref{fig:DT_combined}, the modeled velocity of the planetary shadow is marked by the dashed line to aid the eye.}
    \label{fig:harpsDT}
\end{figure}

Line broadening kernels from the HARPS observations were derived via a least-squares deconvolution, similar to that performed on the MIKE observations. The spectra were deconvolved against a set of ATLAS9 non-rotating spectral templates over the spectral range of $3980-5490$\,\AA, yielding a series of line broadening kernels. Because HARPS is a highly-stablized fiber-fed spectrograph, the line profiles were free of systematic variations. The line profile variations measured by the combined HARPS observations are shown in Figure~\ref{fig:harpsDT}. 

\subsubsection{Measuring line broadening velocities}
\label{sec:linebroad}

An accurate measurement of the line broadening velocities help us constrain the spectroscopic transit model, and in turn yields a more accurate measurement of the spin-orbit angle of the planet. We consider the effects of rotational broadening, macroturbulence, and instrumental broadening in our fitting of the line profiles. We adopt the analytical prescriptions for rotational line broadening and radial-tangential macroturbulence as per \citet{2005oasp.book.....G}. The broadening kernel is derived from an integration over the stellar disk, since radial-tangential macroturbulence is spatially dependent. As per \citet{2005oasp.book.....G}, we assume a fixed linear limb darkening coefficient of 0.5572 for the stellar disk, interpolated from tables in \citet{2011A&A...529A..75C} to the near white light \emph{Kepler} band. Since the disk integration is computationally intensive, we compute a grid of line broadening profiles spanning $v\sin i_\mathrm{rot}$ values of $2$ to $100\,\mathrm{km\,s}^{-1}$ at steps of $1\,\mathrm{km\,s}^{-1}$ and macroturbulence values of $1$ to $20\,\mathrm{km\,s}^{-1}$ at steps of $0.5\,\mathrm{km\,s}^{-1}$. At each step in the fit minization, we interpolate within the grid to the desired rotational and macroturbulent broadening values, and then convolve the interpolated line broadening kernel with a Gaussian kernel representing the instrument profile.

We fit for the macroturbulence and rotational broadening values of each out of transit spectrum obtained with MIKE, TRES, and HARPS. The MIKE out-of-transit observations yielded a rotational broadening velocity $v\sin i_\mathrm{rot} = 13.22 \pm 0.16\,\mathrm{km\,s}^{-1}$, and a macroturbulent broadening velocity of $v_\mathrm{macro} = 6.8 \pm 1.7\,\mathrm{km\,s}^{-1}$, with the uncertainties measured from the scatter in the measured broadening velocities from exposure to exposure. Similarly, an analysis of 23 existing out of transit TRES observations yielded $v\sin i_\mathrm{rot} = 13.07\pm0.10\,\mathrm{km\,s}^{-1}$ and $v_\mathrm{macro} = 6.20 \pm 0.55\,\mathrm{km\,s}^{-1}$, while the HARPS out-of-transit observations yielded $v\sin i_\mathrm{rot} = 12.945\pm0.036\,\mathrm{km\,s}^{-1}$  and $v_\mathrm{macro} = 6.20 \pm 0.20 \,\mathrm{km\,s}^{-1}$. The rotational and macroturbulent broadening velocities from each instrument agree to within $1\sigma$. However, we still expect that the Gaussian approximation for the instrument profile is not perfect, and systematic uncertainties are not accounted for in this analysis. To inflate our uncertainties, we take the maximum deviation between the derived values from the instruments as our uncertainties, and the mean value as our adopted broadening velocities. Our adopted values for the remainder of our analyses are $v\sin i_\mathrm{rot} = 13.08\pm0.28\,\mathrm{km\,s}^{-1}$ and $v_\mathrm{macro} = 6.40 \pm 0.60\,\mathrm{km\,s}^{-1}$.

\section{Doppler tomographic analysis and modeling}
\label{sec:modelling}

We make use of the individual Doppler tomographic transits and \emph{K2} photometry in our modeling of HD 106315c. To derive a consistent set of system parameters, we employ global fit that simultaneously models the \emph{K2} transit light curves and the Doppler tomographic observations. The transit light curves are described by a \citet{2002ApJ...580L.171M} model. The Doppler tomographic signal is modeled by a 2D integration of the stellar surface occulted by the planet \citep[following ][]{2014ApJ...790...30J,2016AJ....152..136Z}, incorporating variations in the projected velocity, limb darkening, and radial-tangential macroturbulence. 

Our model fits for the orbital period $(P)$, transit centroid $(T_0)$, planet-star radius ratio $(R_p/R_\star)$, normalized semi-major axis $(a/R_\star)$, transit inclination $(inc)$, projected spin-orbit angle $(\lambda)$, and rotational broadening $(v\sin i_\mathrm{rot})$. We also fit for the broadening of the planetary shadow due to the macroturbulence $v_\mathrm{marcro}$ on the stellar surface under the projected shadow of the planet in our transit model.  We also incorporate the Dartmouth evolutionary tracks \citep{2008ApJS..178...89D}, spectroscopic stellar parameters for effective temperature $(T_\mathrm{eff})$, metallicity ([Fe/H]), and the \emph{Gaia} DR2 parallax \citep{2018arXiv180409365G} to constrain the system parameters simultaneous to the model fitting \citep[as per][]{Seager:2003,Sozzetti:2007}. Gaussian priors are enforced on the \teff{}, \logg{}, \feh{}, $v\sin i_\mathrm{rot}$, and $v_\mathrm{macro}$ parameters based on their spectroscopic properties from \citet{2017AJ....153..256R}. We also apply Gaussian priors on the parameters $P$ and $T_0$ based on the analyses of \citet{2017A&A...603L...5L}, who observed an additional ground-based transit to refine the ephemeris of the system. Uniform priors are applied to all other free parameters. We hold fixed the quadratic stellar limb darkening parameters, interpolated from the tables in \citet{2011A&A...529A..75C} to the respective stellar properties and photometric filters. Quadratic stellar limb darkening parameters are also assumed in the modeling of the Doppler tomographic signal, with the bandpass assumed to be equivalent to that of the \emph{Kepler} band. The best fit parameters and posteriors are explored through a Markov chain Monte Carlo routine via the affine invariant ensemble sampler \textsc{emcee} \citep{2013PASP..125..306F}, and are tabulated in Table~\ref{tab:parameters}. In addition, we also report the results of fits using the Doppler tomographic observations from each independent instrument in Table~\ref{tab:parameters}, in all cases the spectroscopic observations from each instrument yield a low obliquity orbit for HD 106315c.

\begin{table*}
  \caption{\label{tab:parameters}Global modelling parameters}
  \scalebox{0.9}{

{\renewcommand{\arraystretch}{2.0}
  \begin{tabular}{lrrrrr}
    \hline\hline
    Parameter & \textbf{Global} & MIKE/Magellan & TRES & HARPS & Priors \\
    \hline
    \multicolumn{6}{l}{\emph{Fitted parameters}} \\
    Period (days) & $\mathbf{21.05603_{-0.00026}^{+0.00022}}$ & $21.05607_{-0.00029}^{+0.00031}$ & $21.05697_{-0.00051}^{+0.00048}$ & $21.05689_{-0.00045}^{+0.00037}$ & $\mathcal{G}(21.05683,0.00053)$ \\
    $T_0$ (BJD)   & $\mathbf{2457611.13292 \pm 0.00087}$ & $2457611.13317_{-0.00085}^{+0.00089}$ & $2457611.13295_{-0.00083}^{+0.00090}$ & $2457611.13293_{-0.00092}^{+0.00091}$ & $\mathcal{G}(2457611.1313,0.0045)$\\
    $R_p/R_\star$  & $\mathbf{0.03266 \pm 0.00032}$ & $0.03265 \pm 0.00033$ & $0.032618_{-0.00032}^{+0.00036}$ & $0.03260_{-0.00033}^{+0.00036}$ & $\mathcal{U}$\\
    $a/R_\star$  & $\mathbf{24.79_{-0.39}^{+0.43}}$ & $24.89_{-0.41}^{+0.42}$ & $24.99_{-0.44}^{+0.40}$ & $25.06_{-0.45}^{+0.43}$ & $\mathcal{U}$\\
    $inc \,(^\circ)$  & $\mathbf{88.304_{-0.050}^{+0.057}}$ & $88.323_{-0.054}^{+0.056}$ & $88.337_{-0.057}^{+0.055}$ & $88.346_{-0.062}^{+0.056}$ & $\mathcal{U}$\\
    $\lambda\,(^\circ)$  & $\mathbf{\thislambda}$ & $-15.7_{-4.9}^{+4.8}$ & $21\pm21$ & $11.8\pm4.9$ & $\mathcal{U}$\\
    $v\sin i_\mathrm{rot}\,(\text{km\,s}^{-1})$  & $\mathbf{13.00 \pm 0.28}$ & $12.57\pm0.56$ & $13.08\pm0.29$ & $13.15_{-0.28}^{+0.28}$ & $\mathcal{G}(13.08,0.28)$ \\
    $v_\mathrm{macro}\,(\text{km\,s}^{-1})$  & $\mathbf{7.07_{-0.50}^{+0.51}}$ & $6.48_{-0.56}^{+0.54}$ & $6.41_{-0.63}^{+0.61}$ & $6.73_{-0.51}^{+0.56}$ & $\mathcal{G}(6.40,0.60)$ \\
    $T_\mathrm{eff}$ (K)  & $\mathbf{6244_{-51}^{+49}}$ & $6246_{-52}^{+50}$ & $6253_{-53}^{+55}$ & $6258_{-51}^{+50}$ & $\mathcal{G}(6251,52)$ \\
    $\mathrm{[Fe/H]}$ (dex) & $\mathbf{-0.27\pm0.079}$ & $-0.28_{-0.076}^{+0.079}$ & $-0.272 \pm 0.083$ & $-0.259\pm0.080$ & $\mathcal{G}(-0.27,0.08)$ \\

    \multicolumn{6}{l}{\emph{Inferred parameters}} \\
    $M_\star\,(M_\odot)$ & $\mathbf{1.024_{-0.037}^{+0.035}}$ & $1.023_{-0.036}^{+0.037}$ & $1.027_{-0.034}^{+0.037}$ & $1.031_{-0.035}^{+0.036}$ & -\\
    $R_\star\,(R_\odot)$ & $\mathbf{1.299_{-0.025}^{+0.029}}$ & $1.292_{-0.027}^{+0.032}$ & $1.292_{-0.029}^{+0.031}$ & $1.294 \pm 0.029$ & -\\
    $R_p \, (R_\mathrm{Earth})$ & $\mathbf{4.786\pm0.090}$  & $4.777\pm 0.096$ & $4.764\pm0.089$ & $4.760\pm0.092$ & -\\
    \hline
  \end{tabular}}}
\end{table*}

\begin{figure}
    \centering
    \includegraphics[width=6cm]{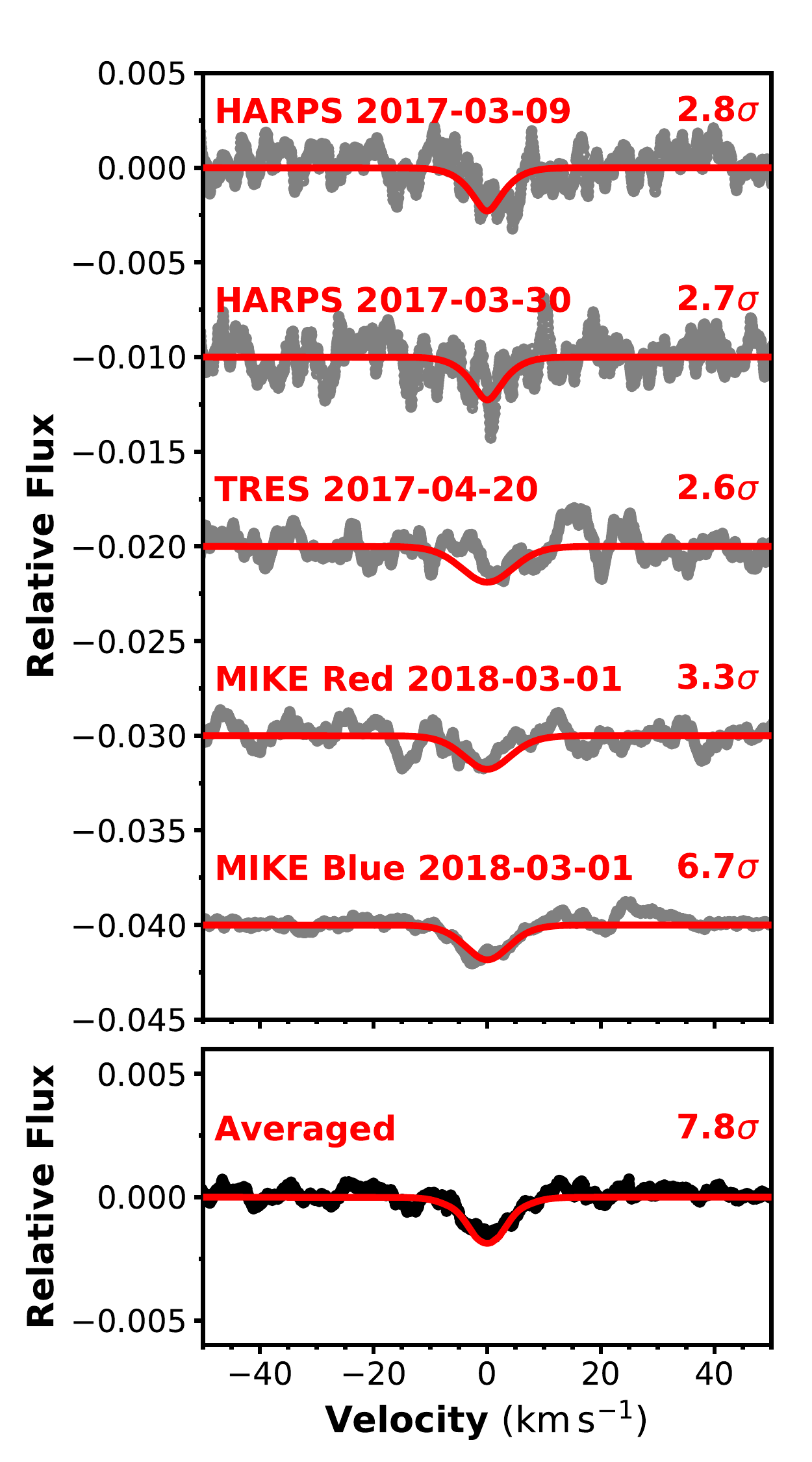}
    \caption{The planetary shadow is summed along its modeled track and plotted to help the reader assess the detection significance of each dataset. The \textbf{top panel} shows the summed Doppler tomographic profile over each individual transit. The \textbf{bottom panel} shows the planetary shadow combined over all observations. The red line shows a Gaussian model fit to the planetary shadow. The detection significance is labelled for each dataset. }
    \label{fig:snr}
\end{figure}

We report an obliquity measurement of $\lambda = \thislambda{} ^\circ$ for HD 106315c from our Doppler tomographic observations, consistent with a well aligned orbit. The significance of the Doppler tomographic detection can be estimated by aligning the planetary shadow during each transit along its predicted velocity based on our best fit model. Figure~\ref{fig:snr} shows the averaged planetary shadow for each of the observations, and for the combined dataset. We find that the planetary shadow is detected at a significance of $7.8\,\sigma$ over the four spectroscopic transits. In fact, the Doppler tomographic signal is detected in each dataset, at $2.8\,\sigma$ and $2.7\,\sigma$ over the HARPS observations on 2017 Mar 09 and 2017 Mar 30, at $2.6\,\sigma$ with the TRES observation on 2017 Apr 20, at $3.3\,\sigma$ with the red arm of the MIKE/Magellan observation on 2018 Mar 01, and at $6.7\,\sigma$ using the observations from the blue arm of MIKE/Magellan.

\section{Discussion}
\label{sec:discussion}

\subsection{Dynamical constraints on the system mutual inclination}
\label{sec:nbody}

Though both the in-situ and in-disk migration scenarios predict well aligned planetary systems with low mutual inclinations, it is possible to excite the mutual inclinations of compact systems via exterior Jovian planets \citep[e.g.][]{2017MNRAS.468.3000M,2017AJ....153..210H,2017MNRAS.468..549B}. Our Doppler tomographic observations constrain the spin-orbit angle of HD 106315c, the largest planet in the system. The inner planet, HD 106315b, is too small for spectroscopic transit follow-up by the current generation of instruments. Despite this, we can still constrain the obliquity and mutual inclination of the system via dynamical arguments. 

It is difficult for a system of Neptunes with a large mutual inclination to maintain a constant transit node. We setup a series of N-body models of the HD 106315 system with the \emph{REBOUND} tool \citep{2012A&A...537A.128R} to investigate the impact of mutual inclination on the transit configuration. We initiate 5000 models of the system, starting with configurations where both planets transit, and measure the fraction of times the systems remain in this alignment given an initial mutual inclination. The models are initiated with the two planets having a uniform distribution of mutual inclinations. The semi-major axes and masses for the model systems are drawn from Gaussian distributions about their published values from \citet{2017A&A...608A..25B}. Given the lack of a strict constraint on the eccentricity of the two planets, we draw eccentricities from uniform distributions based on the $1\sigma$ upper limits given in \citet{2017A&A...608A..25B}, with $\mathcal{U}(0,0.2)$ for planet b, and $\mathcal{U}(0,0.37)$ for planet c, with the argument of periapsis $\omega$ drawn from a uniform distribution of $\mathcal{U}(0,2\pi)$. 

\begin{figure}
    \centering
    \includegraphics[width=7cm]{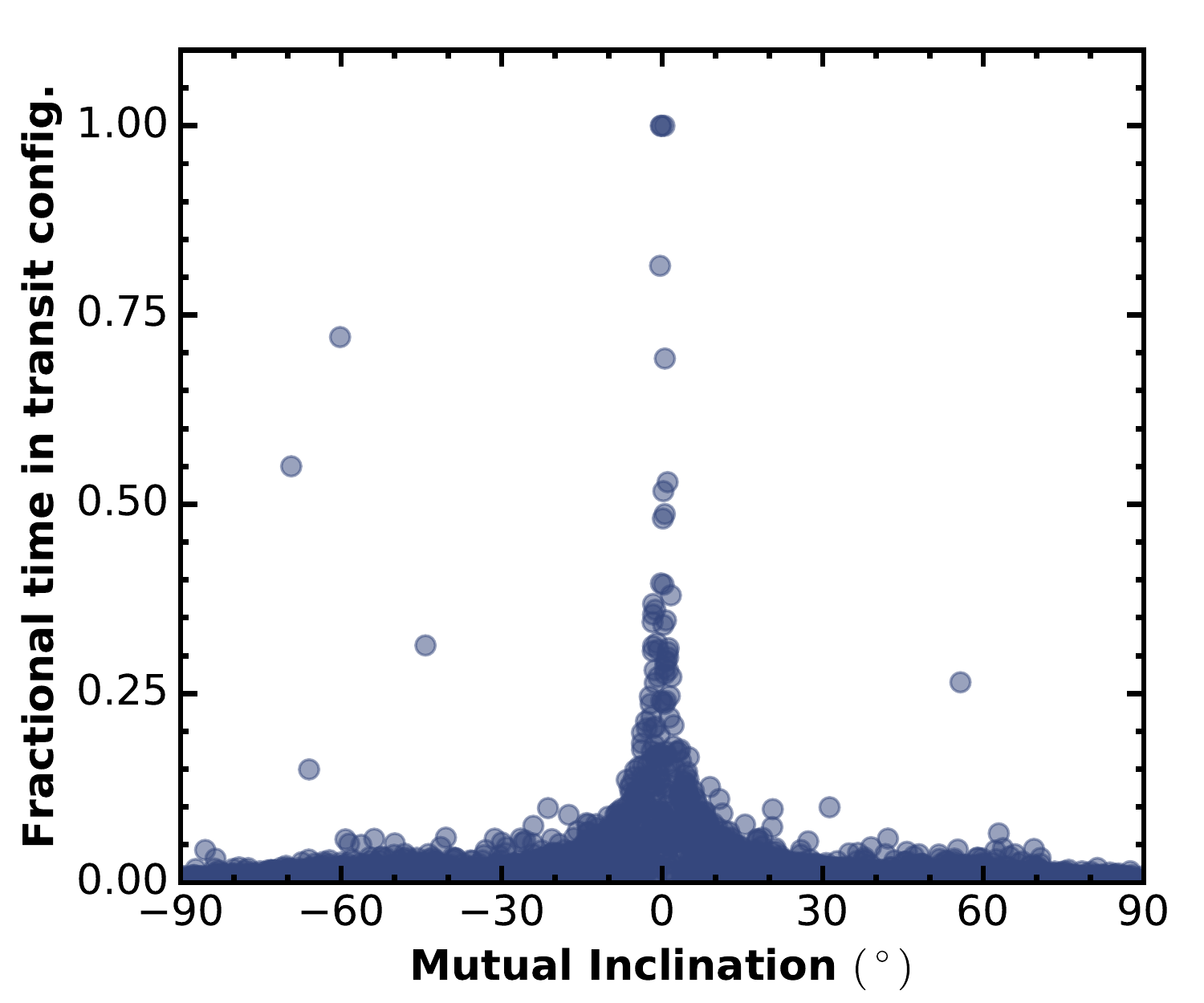}
    \caption{N-body simulations constrain the mutual inclination of the planets in HD 106315. We plot here the fraction of times both planets transit in our modeled systems as a function of the initial mutual inclination of the two planets. The simulation involves 5000 systems with uniformly distributed initial mutual inclinations, integrated over $10^5$ years, only those with inclinations $<5^\circ$ exhibit transits for $>10\%$ of the simulation.}
    \label{fig:mutualinc}
\end{figure}

We allow the N-body models to run for $10^5$ years, and calculate the fraction of times both planets remain in the transit configuration from our line of sight. Figure~\ref{fig:mutualinc} shows the timescale of this dual-transit configuration as a function of the initial mutual inclination of the system. Only those systems with initial mutual inclinations $< 5^\circ$ remain in the dual-transit configuration for more than 10\% of the time. Systems with larger initial mutual inclinations tend to precess out of the dual-transit geometry within $\sim 100$ years. Given the rapid fall off of the dual-transit timescale at higher mutual inclinations, and the low a-priori geometric likelihood of observing dual transits in a high mutual inclination system, we conclude that the orbit normals of both planets in HD 106315 are well aligned with the spin axis of the host star. However, if the system does have a significant mutual inclination, we expect to detect precession of the orbits via transit duration variations, and eventually, the disappearance of transits, within the decades timescale. 

We note that our N-body model assumes a three body system. The possibility of a third outer planet was noted in \citet{2017AJ....153..255C}, though \citet{2017A&A...608A..25B} saw no evidence for additional planets in the HARPS radial velocities. For an external companion to influence the mutual inclination of planets b and c, it needs to be close-in enough to break the coupling between the inner planets. The coupling of planets b and c under the influence of an external companion can be estimated as per \citet[][Equation 12]{2017AJ....153...42L}. We estimate that planet b and c can be considered coupled ($\epsilon_{12} \ll 1$) for external planets as massive as $15\,M_\mathrm{Earth}$ at semi-major axes beyond 0.4 AU (92 day period). Saturn massed companions will need to be beyond 0.75 AU for the inner planets to remain coupled.

\subsection{The obliquities of warm Neptune systems}
\label{sec:discuss_systems}

HD 106315 is an F-star hosting a system of two planets, a super-Earth in a 9.55\,day orbit, and a Neptune in a 21.06\,day orbit. We presented a series of Doppler tomographic observations for the transits of the warm Neptune HD 106315c, finding its projected spin-orbit axis to be well aligned with the spin axis of the host star. From these observations, and dynamical arguments presented in  Section~\ref{sec:nbody}, we find the planets in the HD 106315 system likely inhabit low obliquity, low mutual inclination orbits.  

\begin{figure*}
    \centering
    \begin{tabular}{cc}
         \includegraphics[height=6.5cm]{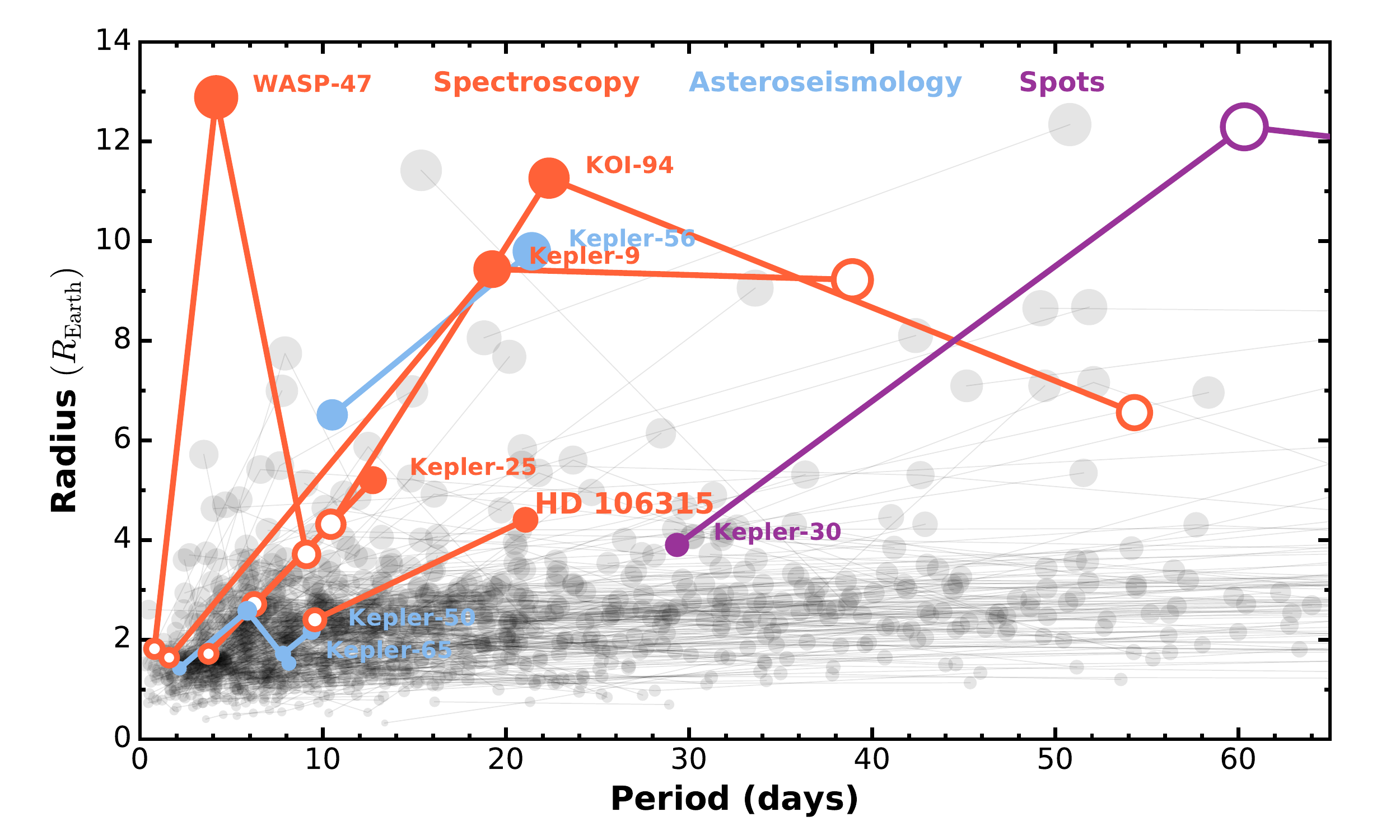} & \includegraphics[height=6.5cm]{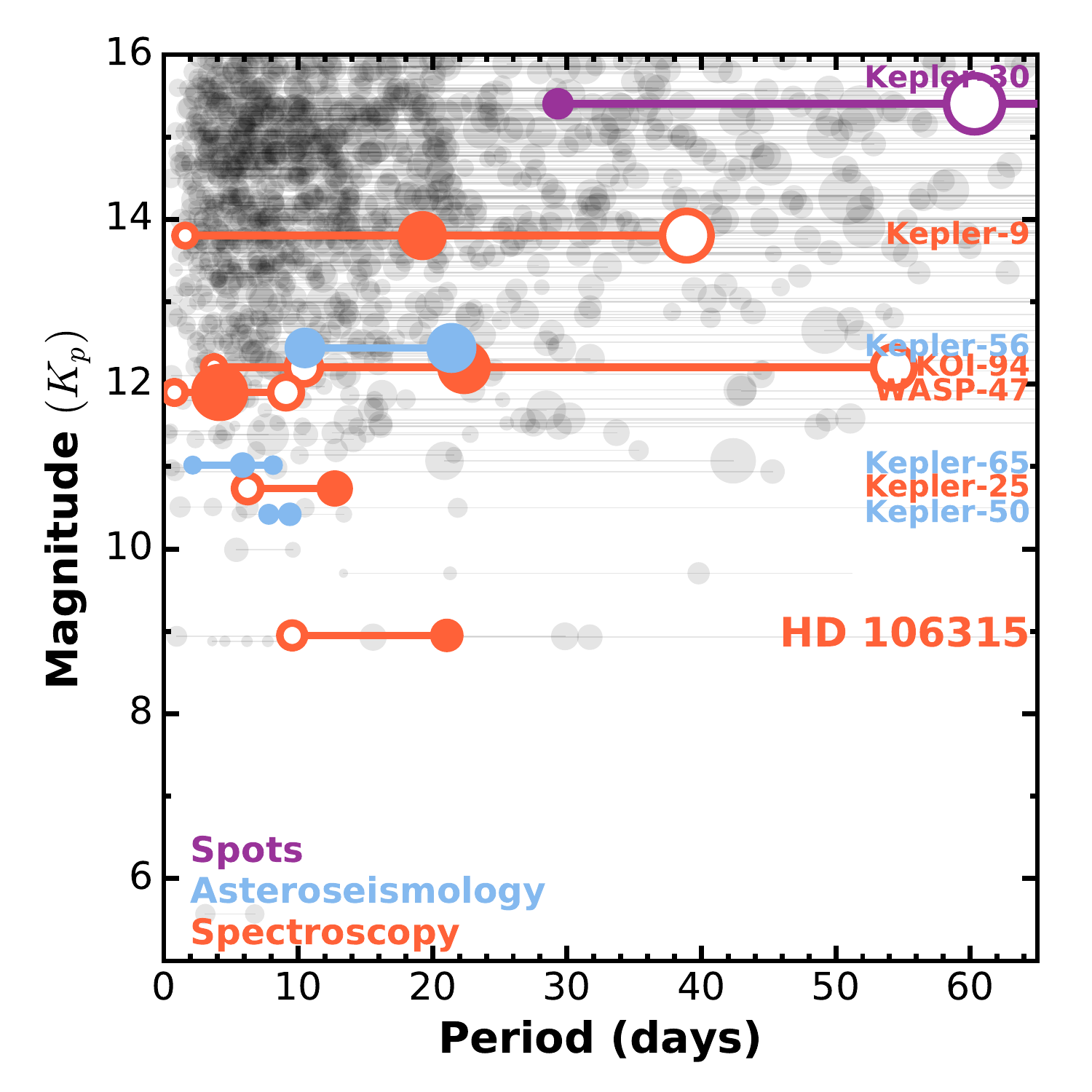}\\
    \end{tabular}
    \caption{Distributions of transiting multi-planet systems in period, planet radius, and visual magnitudes. Planetary systems with obliquities measured spectroscopically are plotted in orange, asteroseismology in blue, and spot crossing analyses in purple. Note that analyses of Kepler-25 involved both spectroscopic transits \citep{2012Natur.487..449S} and asteroseismology results \citep{2014PASJ...66...94B,2016ApJ...819...85C}. Planets for which obliquities were measured are plotted with filled circles, those in the same system but without obliquities directly measured are marked by open circles, and linked by lines. Known multi-planet transiting systems without obliquity measurements are plotted in grey, planets in the same system are linked by lines. The sizes of the points correspond to the radii of the planets. Note that HD 106315c is among the smallest planets to have its obliquities measured spectroscopically, enabled by the brightness of the host star.
    }
    \label{fig:distribution}
\end{figure*}

HD 106315 is only the ninth multi-planet transiting system to have the obliquity of at least one planet measured. Figure~\ref{fig:distribution} shows these nine system in the context of all multi-planet transiting systems. Spectroscopic transits and analyses of spot crossing events have been performed on a number of Jovian-sized planets in the multi-planet systems, including KOI-94 \citep{2012ApJ...759L..36H,2013ApJ...771...11A}, Kepler-30 \citep{2012Natur.487..449S}, WASP-47 \citep{2015ApJ...812L..11S}, and Kepler-9 \citep{2018AJ....155...70W}, finding these to be well aligned systems. The warm Neptune systems Kepler-25 \citep{2013ApJ...771...11A,2014PASJ...66...94B,2016ApJ...819...85C}, Kepler-50, Kepler-60 \citep{2013ApJ...766..101C}, and HD 106315 all inhabit low obliquity orbits too. The Kepler-56 system, with two warm Jupiters, is the only system found to be misaligned via asteroseismology \citep{2013Sci...342..331H}.

In the hot Jupiter obliquity distribution, planets orbiting stars cooler than the Kraft break \citep{1967ApJ...150..551K,1970saac.book..385K} tend to be found in low obliquity orbits, whilst those orbiting hotter stars exhibit a wide distribution of angles \citep[e.g.][]{2010ApJ...718L.145W,2012ApJ...757...18A}. Numerous mechanism have been posed to explain this trend. Tidal interactions between the convective envelope of cool stars and the strong gravitational pull of close-in massive planets can re-align the spin axis of the outer envelope of the host star \citep[e.g.][]{2012MNRAS.423..486L,2013ApJ...769L..10R,2014ApJ...784...66X}. Torquing of the inner disk by massive exterior inclined companions may also occur \citep[e.g.][]{2012Natur.491..418B,2014MNRAS.440.3532L}, and realignment of the disk of cooler disks around cooler stars may happen more quickly than those around hotter stars, leading to the observed distribution \citep{2017arXiv171207655Z}. 

\begin{figure}
    \centering
    \includegraphics[height=8cm]{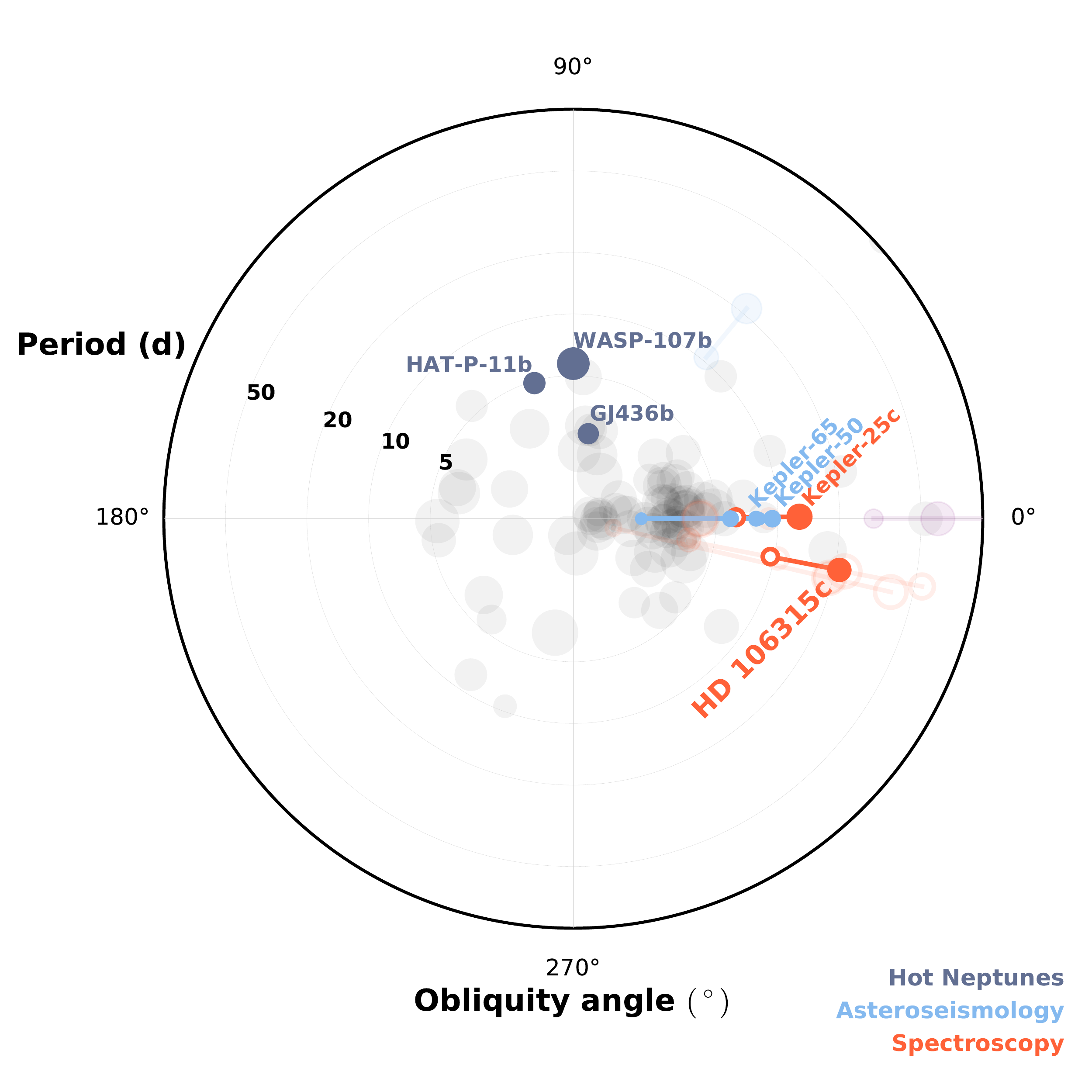}
    \caption{The obliquity distribution of planets as a function of orbital period. The obliquity $\lambda$ is plotted along the polar axis, period along the radial axis scaled logarithmically. The sizes of the points correspond with the radius of the planets, as per left panel of Figure~\ref{fig:distribution}. Planets of Neptune radius or mass are coloured in this figure, with systems for which projected obliquities were measured spectroscopically in orange, and the line-of-sight stellar inclination measured by asteroseismology in blue. Single hot Neptunes with obliquities measured are shown in dark blue. Systems involving Jovian sized planets with obliquities measured are plotted under the same color scheme, but at lower opacity. Other single-planet systems for which obliquities have been measured are plotted in the background at lower opacity. We note that \citet{2017AJ....153..205D} determined WASP-107b to be in a misaligned orbit, but the allowed obliquity range is large $(40-140^\circ)$. The system parameters are retrieved from the TEPCAT Rossiter-McLaughlin Encyclopedia. Figure format inspired from talks by J. Winn. 
    }
    \label{fig:period_lambda}
\end{figure}

In this context, the warm Neptune systems Kepler-25, Kepler-50, Kepler-65, and HD 106315 all orbit F stars close to the Kraft break, and are all found to be in well-aligned orbits. Their obliquities are shown as a function of orbital period in Figure~\ref{fig:period_lambda}. For comparison, in the hot-Jupiter sample 17 planets orbit stars with $6000 < T_\mathrm{eff} < 6300$\,K, of which four (24\%) are found in misaligned geometries. The Neptunes in the multi-planet systems are all too distant, and too low mass, to have influenced the spins of their host stars tidally. In the in-situ formation scenario, we should expect all of these Neptune systems to be found in well aligned orbits barring the torquing of the protoplanetary disk. As such, these warm Neptunes systems around F-stars are perfect laboratories to test the validity of this disk torquing mechanism. It is also interesting to note that the only other transiting Neptunes to have obliquities directly measured, the hot-Neptunes HAT-P-11b \citep{2010ApJ...723L.223W,2011PASJ...63S.531H,2011ApJ...743...61S}, WASP-107b \citep{2017AJ....153..205D}, and GJ 436b \citep{2018Natur.553..477B}, were all found in highly misaligned orbits around cooler stars (Figure~\ref{fig:period_lambda}). In all three cases, the low mass of the planets means the tidal re-alignment timescale is long despite their shorter orbital periods. If this trend holds, we may be seeing a dramatically different early history for hot and warm Neptunes.  

We note that a dichotomy of properties exists between single planet systems and multi-planet systems in eccentricity \citep{2016PNAS..11311431X} and obliquity \citep{2014ApJ...796...47M}. The warm and hot Jupiter populations also differ in multiplicity rates \citep{2012PNAS..109.7982S,2016ApJ...825...98H}: warm Jupiters are more likely to be found with exterior and interior companions than hot Jupiters. The pathways that led to these dichotomies are not yet clear. Hot Jupiters that have dynamically migrated inward are likely to exhibit a variety of obliquities, and may also have disrupted any inner planetary systems \citep{2015ApJ...808...14M}. These hot Jupiters may also have formed in-situ, and then underwent dynamical evolution that torqued the inner hot Jupiter's obliquity to be misaligned to that of the outer planets \citep{2016ApJ...829..114B,2017AJ....154...93S}. Differentiating between these scenarios will require additional radial velocity follow-up of existing single-planet hot Jupiter and Neptune systems \citep[e.g.][]{2018AJ....155..255Y}, or to search for non-transiting inner gas giants to compact multi-planet systems.

\subsection{Future prospects}

The intrinsically small radius ratios of Neptune systems make spectroscopic determination of obliquites difficult, HD 106315c is amongst the smallest planets to have its obliquity directly measured. Systems of Neptunes around bright, relatively rapidly rotating stars like HD 106315 are optimal for such observations. Luckily, we can expect to rapidly build on the obliquities of the four warm Neptune systems and two hot Neptunes in the near future.

With the recent successful launch of the Transiting Exoplanet Survey Satellite \citep[TESS,][]{2016SPIE.9904E..2BR}, we will soon see thousands of new planets to be found around bright, hot stars. We expect TESS to find 35 multiple transiting planetary systems orbiting stars brighter than $T_\mathrm{mag} < 10$ and hotter than $T_\mathrm{eff} > 6500\,\mathrm{K}$, totalling 77 planets in multi-planet systems during its first year of operations (Huang et al. in-prep). Of these, 39 planets in these transiting multi-planets will be of Neptune size $(2 < Rp < 6 \, R_\mathrm{Earth})$ and suitable for further follow-up. In addition, we expect 217 Neptunes in single planet systems around similarly bright stars from the first year of TESS observations. A comparative sample of these populations will help us map the formation pathways of close-in Neptune-sized planets. 

\begin{figure*}
    \centering
    \includegraphics[height=14cm]{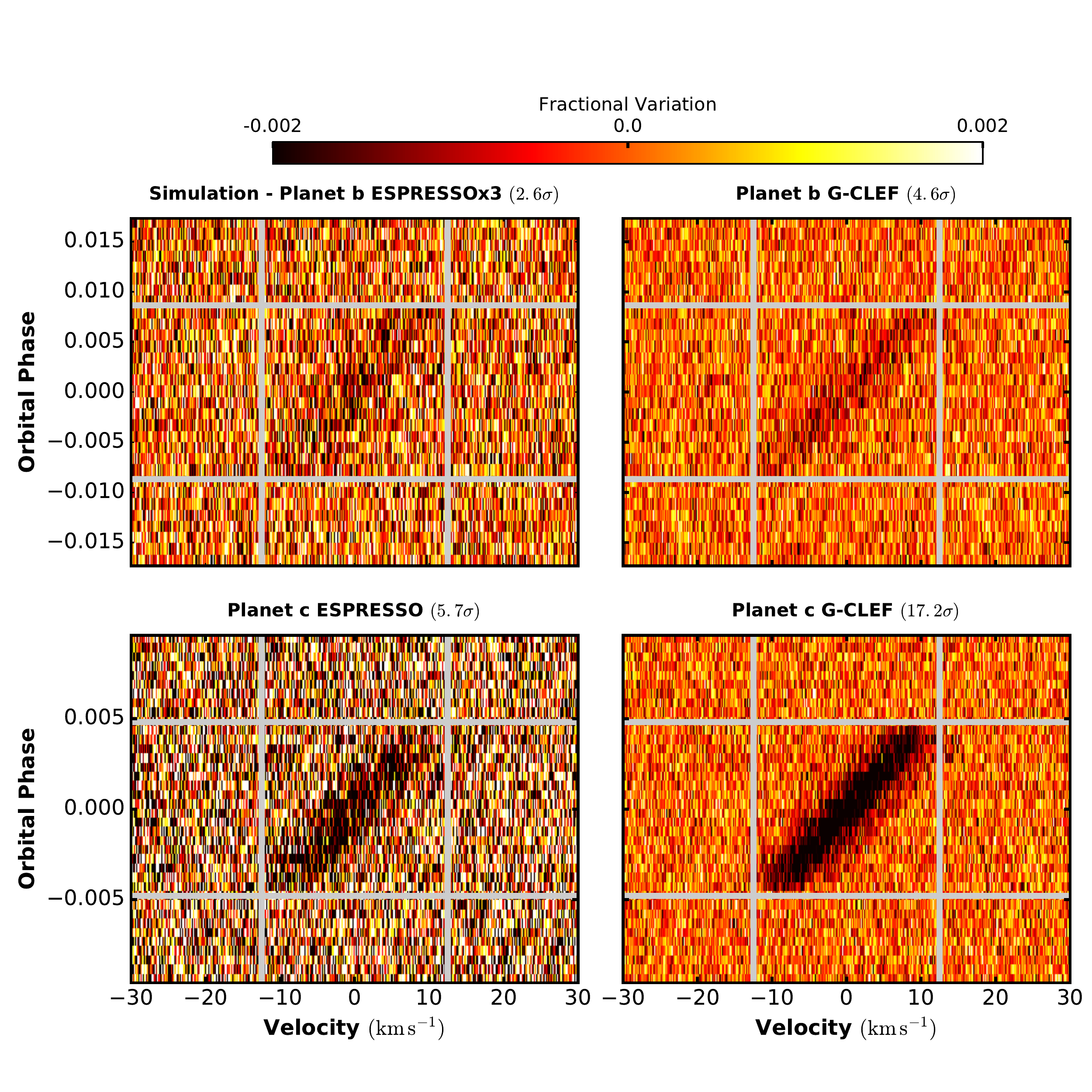}  
    \caption{Simulated Doppler tomographic signals of the HD 106315 system with the next generations of spectrographs. The \textbf{top panels} show simulated signals of the super Earth HD 106315b as observed from three transits from ESPRESSO and a single transit with G-CLEF. The \textbf{bottom panels} show the simulated signals for the warm Neptune HD 106315c, as observed from single transits with ESPRESSO and G-CLEF. The detection significance is indicated on the titles of each sub-panel.}
    \label{fig:DT_simulation}
\end{figure*}

As we have shown, stable instrument profiles are essential in easily tracing the spectroscopic transits of smaller planets. Luckily, a new generation of stable, highly precise spectrographs on large telescopes are coming online. Figure~\ref{fig:DT_simulation} shows the expected Doppler tomographic signal of planets in the HD 106315 system as measured by ESPRESSO utilising one unit of the 8\,m Very Large Telescopes \citep{2014AN....335....8P}, and G-CLEF on the 24\,m  Giant Magellan Telescope \citep{2012SPIE.8446E..1HS}. We expect the Doppler tomographic signal of Neptune-sized planets around stars brighter than $V_\mathrm{mag} = 10$ to be detectable with a single transit from ESPRESSO at the $5\sigma$ significance, while $\sim 2\,R_\mathrm{Earth}$ super Earths like HD 106315b will require spectrographs on the upcoming generation of giant segmented mirror telescopes for clear detections from single transits. The detectability of the spectroscopic transits increases linearly with rotation of the host star. For a system around an equivalent host star rotating at $v\sin i_\mathrm{rot} = 25\,\mathrm{km\,s}^{-1}$, we expect a $4.7\,\sigma$ detection of planet b from three ESPRESSO transits.

\acknowledgements  
We thank the referee for their comments and suggestions, which have improved the quality of this manuscript substantially. 
We thank Jessica Mink for running the TRES pipeline and maintaining the TRES archive. We acknowledge Andrew H. Szentgyorgyi, Gabor F\H{u}r\'{e}sz, and John Geary, who played major roles in the development of the TRES instrument.
The modelling in this paper was performed on the Smithsonian Institution High Performance Cluster (SI/HPC). 
Simulations in this paper made use of the REBOUND code which can be downloaded freely at \url{http://github.com/hannorein/rebound}.
Work by G.Z. is provided by NASA through Hubble Fellowship grant HST-HF2-51402.001-A awarded by the Space Telescope Science Institute, which is operated by the Association of Universities for Research in Astronomy, Inc., for NASA, under contract NAS 5-26555.
Work performed by J.E.R. was supported by the Harvard Future Faculty Leaders Postdoctoral fellowship.
Work by A.V is performed in part under contract with the California Institute of Technology/Jet Propulsion Laboratory funded by NASA through the Sagan Fellowship Program executed by the NASA Exoplanet Science Institute. 
Work by C.H. is supported by the Juan Carlos Torres Fellowship.
Facilities: 
\facility{Magellan:Clay (MIKE), FLWO:1.5m (TRES), ESO:3.6m (HARPS)}

\bibliographystyle{apj}
\bibliography{mybibfile}

\begin{thebibliography}{78}
\expandafter\ifx\csname natexlab\endcsname\relax\def\natexlab#1{#1}\fi

\bibitem[{{Albrecht} {et~al.}(2013){Albrecht}, {Winn}, {Marcy}, {Howard},
  {Isaacson}, \& {Johnson}}]{2013ApJ...771...11A}
{Albrecht}, S., {Winn}, J.~N., {Marcy}, G.~W., {et~al.} 2013, \apj, 771, 11

\bibitem[{{Albrecht} {et~al.}(2012){Albrecht}, {Winn}, {Johnson}, {Howard},
  {Marcy}, {Butler}, {Arriagada}, {Crane}, {Shectman}, {Thompson}, {Hirano},
  {Bakos}, \& {Hartman}}]{2012ApJ...757...18A}
{Albrecht}, S., {Winn}, J.~N., {Johnson}, J.~A., {et~al.} 2012, \apj, 757, 18

\bibitem[{{Bakos} {et~al.}(2010){Bakos}, {Torres}, {P{\'a}l}, {Hartman},
  {Kov{\'a}cs}, {Noyes}, {Latham}, {Sasselov}, {Sip{\H o}cz}, {Esquerdo},
  {Fischer}, {Johnson}, {Marcy}, {Butler}, {Isaacson}, {Howard}, {Vogt},
  {Kov{\'a}cs}, {Fernandez}, {Mo{\'o}r}, {Stefanik}, {L{\'a}z{\'a}r}, {Papp},
  \& {S{\'a}ri}}]{2010ApJ...710.1724B}
{Bakos}, G.~{\'A}., {Torres}, G., {P{\'a}l}, A., {et~al.} 2010, \apj, 710, 1724

\bibitem[{{Barros} {et~al.}(2017){Barros}, {Gosselin}, {Lillo-Box}, {Bayliss},
  {Delgado Mena}, {Brugger}, {Santerne}, {Armstrong}, {Adibekyan}, {Armstrong},
  {Barrado}, {Bento}, {Boisse}, {Bonomo}, {Bouchy}, {Brown}, {Cochran},
  {Collier Cameron}, {Deleuil}, {Demangeon}, {D{\'{\i}}az}, {Doyle},
  {Dumusque}, {Ehrenreich}, {Espinoza}, {Faedi}, {Faria}, {Figueira}, {Foxell},
  {H{\'e}brard}, {Hojjatpanah}, {Jackman}, {Lendl}, {Ligi}, {Lovis}, {Melo},
  {Mousis}, {Neal}, {Osborn}, {Pollacco}, {Santos}, {Sefako}, {Shporer},
  {Sousa}, {Triaud}, {Udry}, {Vigan}, \& {Wyttenbach}}]{2017A&A...608A..25B}
{Barros}, S.~C.~C., {Gosselin}, H., {Lillo-Box}, J., {et~al.} 2017, \aap, 608,
  A25

\bibitem[{{Batygin}(2012)}]{2012Natur.491..418B}
{Batygin}, K. 2012, \nat, 491, 418

\bibitem[{{Batygin} {et~al.}(2016){Batygin}, {Bodenheimer}, \&
  {Laughlin}}]{2016ApJ...829..114B}
{Batygin}, K., {Bodenheimer}, P.~H., \& {Laughlin}, G.~P. 2016, \apj, 829, 114

\bibitem[{{Batygin} \& {Laughlin}(2015)}]{2015PNAS..112.4214B}
{Batygin}, K., \& {Laughlin}, G. 2015, Proceedings of the National Academy of
  Science, 112, 4214

\bibitem[{{Becker} \& {Adams}(2017)}]{2017MNRAS.468..549B}
{Becker}, J.~C., \& {Adams}, F.~C. 2017, \mnras, 468, 549

\bibitem[{{Benomar} {et~al.}(2014){Benomar}, {Masuda}, {Shibahashi}, \&
  {Suto}}]{2014PASJ...66...94B}
{Benomar}, O., {Masuda}, K., {Shibahashi}, H., \& {Suto}, Y. 2014, \pasj, 66,
  94

\bibitem[{{Bernstein} {et~al.}(2003){Bernstein}, {Shectman}, {Gunnels},
  {Mochnacki}, \& {Athey}}]{2003SPIE.4841.1694B}
{Bernstein}, R., {Shectman}, S.~A., {Gunnels}, S.~M., {Mochnacki}, S., \&
  {Athey}, A.~E. 2003, in \procspie, Vol. 4841, Instrument Design and
  Performance for Optical/Infrared Ground-based Telescopes, ed. M.~{Iye} \&
  A.~F.~M. {Moorwood}, 1694--1704

\bibitem[{{Boley} {et~al.}(2016){Boley}, {Granados Contreras}, \&
  {Gladman}}]{2016ApJ...817L..17B}
{Boley}, A.~C., {Granados Contreras}, A.~P., \& {Gladman}, B. 2016, \apjl, 817,
  L17

\bibitem[{{Bourrier} {et~al.}(2018){Bourrier}, {Lovis}, {Beust}, {Ehrenreich},
  {Henry}, {Astudillo-Defru}, {Allart}, {Bonfils}, {S{\'e}gransan}, {Delfosse},
  {Cegla}, {Wyttenbach}, {Heng}, {Lavie}, \& {Pepe}}]{2018Natur.553..477B}
{Bourrier}, V., {Lovis}, C., {Beust}, H., {et~al.} 2018, \nat, 553, 477

\bibitem[{{Buchhave} {et~al.}(2010){Buchhave}, {Bakos}, {Hartman}, {Torres},
  {Kov{\'a}cs}, {Latham}, {Noyes}, {Esquerdo}, {Everett}, {Howard}, {Marcy},
  {Fischer}, {Johnson}, {Andersen}, {F{\H u}r{\'e}sz}, {Perumpilly},
  {Sasselov}, {Stefanik}, {B{\'e}ky}, {L{\'a}z{\'a}r}, {Papp}, \&
  {S{\'a}ri}}]{2010ApJ...720.1118B}
{Buchhave}, L.~A., {Bakos}, G.~{\'A}., {Hartman}, J.~D., {et~al.} 2010, \apj,
  720, 1118

\bibitem[{{Campante} {et~al.}(2016){Campante}, {Lund}, {Kuszlewicz}, {Davies},
  {Chaplin}, {Albrecht}, {Winn}, {Bedding}, {Benomar}, {Bossini}, {Handberg},
  {Santos}, {Van Eylen}, {Basu}, {Christensen-Dalsgaard}, {Elsworth}, {Hekker},
  {Hirano}, {Huber}, {Karoff}, {Kjeldsen}, {Lundkvist}, {North}, {Silva
  Aguirre}, {Stello}, \& {White}}]{2016ApJ...819...85C}
{Campante}, T.~L., {Lund}, M.~N., {Kuszlewicz}, J.~S., {et~al.} 2016, \apj,
  819, 85

\bibitem[{{Castelli} \& {Kurucz}(2004)}]{2004astro.ph..5087C}
{Castelli}, F., \& {Kurucz}, R.~L. 2004, ArXiv Astrophysics e-prints

\bibitem[{{Chaplin} {et~al.}(2013){Chaplin}, {Sanchis-Ojeda}, {Campante},
  {Handberg}, {Stello}, {Winn}, {Basu}, {Christensen-Dalsgaard}, {Davies},
  {Metcalfe}, {Buchhave}, {Fischer}, {Bedding}, {Cochran}, {Elsworth},
  {Gilliland}, {Hekker}, {Huber}, {Isaacson}, {Karoff}, {Kawaler}, {Kjeldsen},
  {Latham}, {Lund}, {Lundkvist}, {Marcy}, {Miglio}, {Barclay}, \&
  {Lissauer}}]{2013ApJ...766..101C}
{Chaplin}, W.~J., {Sanchis-Ojeda}, R., {Campante}, T.~L., {et~al.} 2013, \apj,
  766, 101

\bibitem[{{Claret} \& {Bloemen}(2011)}]{2011A&A...529A..75C}
{Claret}, A., \& {Bloemen}, S. 2011, \aap, 529, A75

\bibitem[{{Collier Cameron} {et~al.}(2010{\natexlab{a}}){Collier Cameron},
  {Bruce}, {Miller}, {Triaud}, \& {Queloz}}]{2010MNRAS.403..151C}
{Collier Cameron}, A., {Bruce}, V.~A., {Miller}, G.~R.~M., {Triaud},
  A.~H.~M.~J., \& {Queloz}, D. 2010{\natexlab{a}}, \mnras, 403, 151

\bibitem[{{Collier Cameron} {et~al.}(2010{\natexlab{b}}){Collier Cameron},
  {Guenther}, {Smalley}, {McDonald}, {Hebb}, {Andersen}, {Augusteijn},
  {Barros}, {Brown}, {Cochran}, {Endl}, {Fossey}, {Hartmann}, {Maxted},
  {Pollacco}, {Skillen}, {Telting}, {Waldmann}, \&
  {West}}]{2010MNRAS.407..507C}
{Collier Cameron}, A., {Guenther}, E., {Smalley}, B., {et~al.}
  2010{\natexlab{b}}, \mnras, 407, 507

\bibitem[{{Crossfield} {et~al.}(2017){Crossfield}, {Ciardi}, {Isaacson},
  {Howard}, {Petigura}, {Weiss}, {Fulton}, {Sinukoff}, {Schlieder}, {Mawet},
  {Ruane}, {de Pater}, {de Kleer}, {Davies}, {Christiansen}, {Dressing},
  {Hirsch}, {Benneke}, {Crepp}, {Kosiarek}, {Livingston}, {Gonzales},
  {Beichman}, \& {Knutson}}]{2017AJ....153..255C}
{Crossfield}, I.~J.~M., {Ciardi}, D.~R., {Isaacson}, H., {et~al.} 2017, \aj,
  153, 255

\bibitem[{{Dai} \& {Winn}(2017)}]{2017AJ....153..205D}
{Dai}, F., \& {Winn}, J.~N. 2017, \aj, 153, 205

\bibitem[{{Donati} {et~al.}(1997){Donati}, {Semel}, {Carter}, {Rees}, \&
  {Collier Cameron}}]{1997MNRAS.291..658D}
{Donati}, J.-F., {Semel}, M., {Carter}, B.~D., {Rees}, D.~E., \& {Collier
  Cameron}, A. 1997, \mnras, 291, 658

\bibitem[{{Dong} \& {Zhu}(2013)}]{2013ApJ...778...53D}
{Dong}, S., \& {Zhu}, Z. 2013, \apj, 778, 53

\bibitem[{{Dotter} {et~al.}(2008){Dotter}, {Chaboyer}, {Jevremovi{\'c}},
  {Kostov}, {Baron}, \& {Ferguson}}]{2008ApJS..178...89D}
{Dotter}, A., {Chaboyer}, B., {Jevremovi{\'c}}, D., {et~al.} 2008, \apjs, 178,
  89

\bibitem[{{Foreman-Mackey} {et~al.}(2013){Foreman-Mackey}, {Hogg}, {Lang}, \&
  {Goodman}}]{2013PASP..125..306F}
{Foreman-Mackey}, D., {Hogg}, D.~W., {Lang}, D., \& {Goodman}, J. 2013, \pasp,
  125, 306

\bibitem[{{Fressin} {et~al.}(2013){Fressin}, {Torres}, {Charbonneau}, {Bryson},
  {Christiansen}, {Dressing}, {Jenkins}, {Walkowicz}, \&
  {Batalha}}]{2013ApJ...766...81F}
{Fressin}, F., {Torres}, G., {Charbonneau}, D., {et~al.} 2013, \apj, 766, 81

\bibitem[{Furesz(2008)}]{Furesz2014}
Furesz, G. 2008, PhD thesis, University of Szeged

\bibitem[{{Gaia Collaboration} {et~al.}(2018){Gaia Collaboration}, {Brown},
  {Vallenari}, {Prusti}, {de Bruijne}, {Babusiaux}, \&
  {Bailer-Jones}}]{2018arXiv180409365G}
{Gaia Collaboration}, {Brown}, A.~G.~A., {Vallenari}, A., {et~al.} 2018, ArXiv
  e-prints, 1804.09365

\bibitem[{{Gray}(2005)}]{2005oasp.book.....G}
{Gray}, D.~F. 2005, {The Observation and Analysis of Stellar Photospheres}

\bibitem[{{Gray} \& {Corbally}(1994)}]{1994AJ....107..742G}
{Gray}, R.~O., \& {Corbally}, C.~J. 1994, \aj, 107, 742

\bibitem[{{Griffin}(1973)}]{1973MNRAS.162..243G}
{Griffin}, R. 1973, \mnras, 162, 243

\bibitem[{{Griffin} \& {Griffin}(1973)}]{1973MNRAS.162..255G}
{Griffin}, R., \& {Griffin}, R. 1973, \mnras, 162, 255

\bibitem[{{Hirano} {et~al.}(2011){Hirano}, {Narita}, {Shporer}, {Sato}, {Aoki},
  \& {Tamura}}]{2011PASJ...63S.531H}
{Hirano}, T., {Narita}, N., {Shporer}, A., {et~al.} 2011, \pasj, 63, 531

\bibitem[{{Hirano} {et~al.}(2012){Hirano}, {Narita}, {Sato}, {Takahashi},
  {Masuda}, {Takeda}, {Aoki}, {Tamura}, \& {Suto}}]{2012ApJ...759L..36H}
{Hirano}, T., {Narita}, N., {Sato}, B., {et~al.} 2012, \apjl, 759, L36

\bibitem[{{Huang} {et~al.}(2016){Huang}, {Wu}, \&
  {Triaud}}]{2016ApJ...825...98H}
{Huang}, C., {Wu}, Y., \& {Triaud}, A.~H.~M.~J. 2016, \apj, 825, 98

\bibitem[{{Huang} {et~al.}(2017){Huang}, {Petrovich}, \&
  {Deibert}}]{2017AJ....153..210H}
{Huang}, C.~X., {Petrovich}, C., \& {Deibert}, E. 2017, \aj, 153, 210

\bibitem[{{Huber} {et~al.}(2013){Huber}, {Carter}, {Barbieri}, {Miglio},
  {Deck}, {Fabrycky}, {Montet}, {Buchhave}, {Chaplin}, {Hekker},
  {Montalb{\'a}n}, {Sanchis-Ojeda}, {Basu}, {Bedding}, {Campante},
  {Christensen-Dalsgaard}, {Elsworth}, {Stello}, {Arentoft}, {Ford},
  {Gilliland}, {Handberg}, {Howard}, {Isaacson}, {Johnson}, {Karoff},
  {Kawaler}, {Kjeldsen}, {Latham}, {Lund}, {Lundkvist}, {Marcy}, {Metcalfe},
  {Silva Aguirre}, \& {Winn}}]{2013Sci...342..331H}
{Huber}, D., {Carter}, J.~A., {Barbieri}, M., {et~al.} 2013, Science, 342, 331

\bibitem[{{Johnson} {et~al.}(2014){Johnson}, {Cochran}, {Albrecht},
  {Dodson-Robinson}, {Winn}, \& {Gullikson}}]{2014ApJ...790...30J}
{Johnson}, M.~C., {Cochran}, W.~D., {Albrecht}, S., {et~al.} 2014, \apj, 790,
  30

\bibitem[{{Kelson}(2003)}]{2003PASP..115..688K}
{Kelson}, D.~D. 2003, \pasp, 115, 688

\bibitem[{{Kelson} {et~al.}(2000){Kelson}, {Illingworth}, {van Dokkum}, \&
  {Franx}}]{2000ApJ...531..159K}
{Kelson}, D.~D., {Illingworth}, G.~D., {van Dokkum}, P.~G., \& {Franx}, M.
  2000, \apj, 531, 159

\bibitem[{{Kraft}(1967)}]{1967ApJ...150..551K}
{Kraft}, R.~P. 1967, \apj, 150, 551

\bibitem[{{Kraft}(1970)}]{1970saac.book..385K}
---. 1970, {Stellar Rotation}, ed. G.~H. {Herbig} \& O.~{Struve}, 385

\bibitem[{{Lai}(2012)}]{2012MNRAS.423..486L}
{Lai}, D. 2012, \mnras, 423, 486

\bibitem[{{Lai}(2014)}]{2014MNRAS.440.3532L}
---. 2014, \mnras, 440, 3532

\bibitem[{{Lai} \& {Pu}(2017)}]{2017AJ....153...42L}
{Lai}, D., \& {Pu}, B. 2017, \aj, 153, 42

\bibitem[{{Lee} {et~al.}(2014){Lee}, {Chiang}, \&
  {Ormel}}]{2014ApJ...797...95L}
{Lee}, E.~J., {Chiang}, E., \& {Ormel}, C.~W. 2014, \apj, 797, 95

\bibitem[{{Lendl} {et~al.}(2017){Lendl}, {Ehrenreich}, {Turner}, {Bayliss},
  {Blanco-Cuaresma}, {Giles}, {Bouchy}, {Marmier}, \&
  {Udry}}]{2017A&A...603L...5L}
{Lendl}, M., {Ehrenreich}, D., {Turner}, O.~D., {et~al.} 2017, \aap, 603, L5

\bibitem[{{Mandel} \& {Agol}(2002)}]{2002ApJ...580L.171M}
{Mandel}, K., \& {Agol}, E. 2002, \apjl, 580, L171

\bibitem[{{Mayor} {et~al.}(2003){Mayor}, {Pepe}, {Queloz}, {Bouchy},
  {Rupprecht}, {Lo Curto}, {Avila}, {Benz}, {Bertaux}, {Bonfils}, {Dall},
  {Dekker}, {Delabre}, {Eckert}, {Fleury}, {Gilliotte}, {Gojak}, {Guzman},
  {Kohler}, {Lizon}, {Longinotti}, {Lovis}, {Megevand}, {Pasquini}, {Reyes},
  {Sivan}, {Sosnowska}, {Soto}, {Udry}, {van Kesteren}, {Weber}, \&
  {Weilenmann}}]{2003Msngr.114...20M}
{Mayor}, M., {Pepe}, F., {Queloz}, D., {et~al.} 2003, The Messenger, 114, 20

\bibitem[{{McLaughlin}(1924)}]{1924ApJ....60...22M}
{McLaughlin}, D.~B. 1924, \apj, 60, 22

\bibitem[{{Moehler} {et~al.}(2014){Moehler}, {Modigliani}, {Freudling},
  {Giammichele}, {Gianninas}, {Gonneau}, {Kausch}, {Lan{\c c}on}, {Noll},
  {Rauch}, \& {Vinther}}]{2014A&A...568A...9M}
{Moehler}, S., {Modigliani}, A., {Freudling}, W., {et~al.} 2014, \aap, 568, A9

\bibitem[{{Morton} \& {Winn}(2014)}]{2014ApJ...796...47M}
{Morton}, T.~D., \& {Winn}, J.~N. 2014, \apj, 796, 47

\bibitem[{{Mustill} {et~al.}(2015){Mustill}, {Davies}, \&
  {Johansen}}]{2015ApJ...808...14M}
{Mustill}, A.~J., {Davies}, M.~B., \& {Johansen}, A. 2015, \apj, 808, 14

\bibitem[{{Mustill} {et~al.}(2017){Mustill}, {Davies}, \&
  {Johansen}}]{2017MNRAS.468.3000M}
---. 2017, \mnras, 468, 3000

\bibitem[{{Pepe} {et~al.}(2014){Pepe}, {Molaro}, {Cristiani}, {Rebolo},
  {Santos}, {Dekker}, {M{\'e}gevand}, {Zerbi}, {Cabral}, {Di Marcantonio},
  {Abreu}, {Affolter}, {Aliverti}, {Allende Prieto}, {Amate}, {Avila},
  {Baldini}, {Bristow}, {Broeg}, {Cirami}, {Coelho}, {Conconi}, {Coretti},
  {Cupani}, {D'Odorico}, {De Caprio}, {Delabre}, {Dorn}, {Figueira}, {Fragoso},
  {Galeotta}, {Genolet}, {Gomes}, {Gonz{\'a}lez Hern{\'a}ndez}, {Hughes},
  {Iwert}, {Kerber}, {Landoni}, {Lizon}, {Lovis}, {Maire}, {Mannetta},
  {Martins}, {Monteiro}, {Oliveira}, {Poretti}, {Rasilla}, {Riva}, {Santana
  Tschudi}, {Santos}, {Sosnowska}, {Sousa}, {Span{\'o}}, {Tenegi}, {Toso},
  {Vanzella}, {Viel}, \& {Zapatero Osorio}}]{2014AN....335....8P}
{Pepe}, F., {Molaro}, P., {Cristiani}, S., {et~al.} 2014, Astronomische
  Nachrichten, 335, 8

\bibitem[{{Quinn} {et~al.}(2014){Quinn}, {White}, {Latham}, {Buchhave},
  {Torres}, {Stefanik}, {Berlind}, {Bieryla}, {Calkins}, {Esquerdo}, {F{\H
  u}r{\'e}sz}, {Geary}, \& {Szentgyorgyi}}]{2014ApJ...787...27Q}
{Quinn}, S.~N., {White}, R.~J., {Latham}, D.~W., {et~al.} 2014, \apj, 787, 27

\bibitem[{{Rein} \& {Liu}(2012)}]{2012A&A...537A.128R}
{Rein}, H., \& {Liu}, S.-F. 2012, \aap, 537, A128

\bibitem[{{Ricker} {et~al.}(2016){Ricker}, {Vanderspek}, {Winn}, {Seager},
  {Berta-Thompson}, {Levine}, {Villasenor}, {Latham}, {Charbonneau}, {Holman},
  {Johnson}, {Sasselov}, {Szentgyorgyi}, {Torres}, {Bakos}, {Brown},
  {Christensen-Dalsgaard}, {Kjeldsen}, {Clampin}, {Rinehart}, {Deming}, {Doty},
  {Dunham}, {Ida}, {Kawai}, {Sato}, {Jenkins}, {Lissauer}, {Jernigan},
  {Kaltenegger}, {Laughlin}, {Lin}, {McCullough}, {Narita}, {Pepper},
  {Stassun}, \& {Udry}}]{2016SPIE.9904E..2BR}
{Ricker}, G.~R., {Vanderspek}, R., {Winn}, J., {et~al.} 2016, in \procspie,
  Vol. 9904, Space Telescopes and Instrumentation 2016: Optical, Infrared, and
  Millimeter Wave, 99042B

\bibitem[{{Rodriguez} {et~al.}(2017){Rodriguez}, {Zhou}, {Vanderburg},
  {Eastman}, {Kreidberg}, {Cargile}, {Bieryla}, {Latham}, {Irwin}, {Mayo},
  {Calkins}, {Esquerdo}, \& {Mink}}]{2017AJ....153..256R}
{Rodriguez}, J.~E., {Zhou}, G., {Vanderburg}, A., {et~al.} 2017, \aj, 153, 256

\bibitem[{{Rogers} \& {Lin}(2013)}]{2013ApJ...769L..10R}
{Rogers}, T.~M., \& {Lin}, D.~N.~C. 2013, \apjl, 769, L10

\bibitem[{{Rossiter}(1924)}]{1924ApJ....60...15R}
{Rossiter}, R.~A. 1924, \apj, 60, 15

\bibitem[{{Sanchis-Ojeda} \& {Winn}(2011)}]{2011ApJ...743...61S}
{Sanchis-Ojeda}, R., \& {Winn}, J.~N. 2011, \apj, 743, 61

\bibitem[{{Sanchis-Ojeda} {et~al.}(2012){Sanchis-Ojeda}, {Fabrycky}, {Winn},
  {Barclay}, {Clarke}, {Ford}, {Fortney}, {Geary}, {Holman}, {Howard},
  {Jenkins}, {Koch}, {Lissauer}, {Marcy}, {Mullally}, {Ragozzine}, {Seader},
  {Still}, \& {Thompson}}]{2012Natur.487..449S}
{Sanchis-Ojeda}, R., {Fabrycky}, D.~C., {Winn}, J.~N., {et~al.} 2012, \nat,
  487, 449

\bibitem[{{Sanchis-Ojeda} {et~al.}(2015){Sanchis-Ojeda}, {Winn}, {Dai},
  {Howard}, {Isaacson}, {Marcy}, {Petigura}, {Sinukoff}, {Weiss}, {Albrecht},
  {Hirano}, \& {Rogers}}]{2015ApJ...812L..11S}
{Sanchis-Ojeda}, R., {Winn}, J.~N., {Dai}, F., {et~al.} 2015, \apjl, 812, L11

\bibitem[{{Seager} \& {Mall{\'e}n-Ornelas}(2003)}]{Seager:2003}
{Seager}, S., \& {Mall{\'e}n-Ornelas}, G. 2003, in Astronomical Society of the
  Pacific Conference Series, Vol. 294, Scientific Frontiers in Research on
  Extrasolar Planets, ed. D.~{Deming} \& S.~{Seager}, 419--422

\bibitem[{{Sozzetti} {et~al.}(2007){Sozzetti}, {Torres}, {Charbonneau},
  {Latham}, {Holman}, {Winn}, {Laird}, \& {O'Donovan}}]{Sozzetti:2007}
{Sozzetti}, A., {Torres}, G., {Charbonneau}, D., {et~al.} 2007, \apj, 664, 1190

\bibitem[{{Spalding} \& {Batygin}(2017)}]{2017AJ....154...93S}
{Spalding}, C., \& {Batygin}, K. 2017, \aj, 154, 93

\bibitem[{{Steffen} {et~al.}(2012){Steffen}, {Ragozzine}, {Fabrycky}, {Carter},
  {Ford}, {Holman}, {Rowe}, {Welsh}, {Borucki}, {Boss}, {Ciardi}, \&
  {Quinn}}]{2012PNAS..109.7982S}
{Steffen}, J.~H., {Ragozzine}, D., {Fabrycky}, D.~C., {et~al.} 2012,
  Proceedings of the National Academy of Science, 109, 7982

\bibitem[{{Szentgyorgyi} {et~al.}(2012){Szentgyorgyi}, {Frebel}, {Furesz},
  {Hertz}, {Norton}, {Bean}, {Bergner}, {Crane}, {Evans}, {Evans}, {Gauron},
  {Jord{\'a}n}, {Park}, {Uomoto}, {Barnes}, {Davis}, {Eisenhower}, {Epps},
  {Guzman}, {McCracken}, {Ordway}, {Plummer}, {Podgorski}, \&
  {Weaver}}]{2012SPIE.8446E..1HS}
{Szentgyorgyi}, A., {Frebel}, A., {Furesz}, G., {et~al.} 2012, in \procspie,
  Vol. 8446, Ground-based and Airborne Instrumentation for Astronomy IV, 84461H

\bibitem[{{Wang} {et~al.}(2018){Wang}, {Addison}, {Fischer}, {Brewer},
  {Isaacson}, {Howard}, \& {Laughlin}}]{2018AJ....155...70W}
{Wang}, S., {Addison}, B., {Fischer}, D.~A., {et~al.} 2018, \aj, 155, 70

\bibitem[{{Winn} {et~al.}(2010{\natexlab{a}}){Winn}, {Fabrycky}, {Albrecht}, \&
  {Johnson}}]{2010ApJ...718L.145W}
{Winn}, J.~N., {Fabrycky}, D., {Albrecht}, S., \& {Johnson}, J.~A.
  2010{\natexlab{a}}, \apjl, 718, L145

\bibitem[{{Winn} {et~al.}(2010{\natexlab{b}}){Winn}, {Johnson}, {Howard},
  {Marcy}, {Isaacson}, {Shporer}, {Bakos}, {Hartman}, \&
  {Albrecht}}]{2010ApJ...723L.223W}
{Winn}, J.~N., {Johnson}, J.~A., {Howard}, A.~W., {et~al.} 2010{\natexlab{b}},
  \apjl, 723, L223

\bibitem[{{Xie} {et~al.}(2016){Xie}, {Dong}, {Zhu}, {Huber}, {Zheng}, {De Cat},
  {Fu}, {Liu}, {Luo}, {Wu}, {Zhang}, {Zhang}, {Zhou}, {Cao}, {Hou}, {Wang}, \&
  {Zhang}}]{2016PNAS..11311431X}
{Xie}, J.-W., {Dong}, S., {Zhu}, Z., {et~al.} 2016, Proceedings of the National
  Academy of Science, 113, 11431

\bibitem[{{Xue} {et~al.}(2014){Xue}, {Suto}, {Taruya}, {Hirano}, {Fujii}, \&
  {Masuda}}]{2014ApJ...784...66X}
{Xue}, Y., {Suto}, Y., {Taruya}, A., {et~al.} 2014, \apj, 784, 66

\bibitem[{{Yee} {et~al.}(2018){Yee}, {Petigura}, {Fulton}, {Knutson},
  {Batygin}, {Bakos}, {Hartman}, {Hirsch}, {Howard}, {Isaacson}, {Kosiarek},
  {Sinukoff}, \& {Weiss}}]{2018AJ....155..255Y}
{Yee}, S.~W., {Petigura}, E.~A., {Fulton}, B.~J., {et~al.} 2018, \aj, 155, 255

\bibitem[{{Zanazzi} \& {Lai}(2017)}]{2017arXiv171207655Z}
{Zanazzi}, J.~J., \& {Lai}, D. 2017, ArXiv e-prints, 1712.07655

\bibitem[{{Zhou} {et~al.}(2016{\natexlab{a}}){Zhou}, {Latham}, {Bieryla},
  {Beatty}, {Buchhave}, {Esquerdo}, {Berlind}, \&
  {Calkins}}]{2016MNRAS.460.3376Z}
{Zhou}, G., {Latham}, D.~W., {Bieryla}, A., {et~al.} 2016{\natexlab{a}},
  \mnras, 460, 3376

\bibitem[{{Zhou} {et~al.}(2016{\natexlab{b}}){Zhou}, {Rodriguez}, {Collins},
  {Beatty}, {Oberst}, {Heintz}, {Stassun}, {Latham}, {Kuhn}, {Bieryla}, {Lund},
  {Labadie-Bartz}, {Siverd}, {Stevens}, {Gaudi}, {Pepper}, {Buchhave},
  {Eastman}, {Col{\'o}n}, {Cargile}, {James}, {Gregorio}, {Reed}, {Jensen},
  {Cohen}, {McLeod}, {Tan}, {Zambelli}, {Bayliss}, {Bento}, {Esquerdo},
  {Berlind}, {Calkins}, {Blancato}, {Manner}, {Samulski}, {Stockdale},
  {Nelson}, {Stephens}, {Curtis}, {Kielkopf}, {Fulton}, {DePoy}, {Marshall},
  {Pogge}, {Gould}, {Trueblood}, \& {Trueblood}}]{2016AJ....152..136Z}
{Zhou}, G., {Rodriguez}, J.~E., {Collins}, K.~A., {et~al.} 2016{\natexlab{b}},
  \aj, 152, 136

\end{thebibliography}

\end{document}